\documentclass[modern]{aastex63}
\def\be{\begin{equation}}
\def\ee{\end{equation}}
\def\ba{\begin{eqnarray}}
\def\ea{\end{eqnarray}}
\newcommand{\msun}{\ifmmode\mbox{M}_{\odot}\else$\mbox{M}_{\odot}$\fi}
\newcommand{\rsun}{\ifmmode\mbox{R}_{\odot}\else$\mbox{R}_{\odot}$\fi}
\newcommand{\degrees}{\ifmmode^{\circ}\else$^{\circ}$\fi}
\newcommand{\degree}{\ifmmode^{\circ}\else$^{\circ}$\fi}
\newcommand{\amin}{\ifmmode^{\prime}\else$^{\prime}$\fi}
\newcommand{\asec}{\ifmmode^{\prime\prime}\else$^{\prime\prime}$\fi}

\shorttitle{Timing Solutions of Eight MSPs}
\shortauthors{Deneva et al.}

\begin{document}

%\title{Radio and Gamma-ray Timing of Eight Binary Millisecond Pulsars Found in Arecibo Searches of Fermi Unidentified Sources}

\title{Timing of Eight Binary Millisecond Pulsars Found with Arecibo in Fermi-LAT Unidentified Sources}

\author{J.~S.~Deneva}
\affiliation{George Mason University, resident at the Naval Research Laboratory, Washington, DC 20375, USA}
\author{P.~S.~Ray}
\affiliation{Naval Research Laboratory, Washington, DC 20375, USA}
\author{F.~Camilo}
\affiliation{South African Radio Astronomy Observatory, Cape Town, 7925, South Africa}
\author{P.~C.~C.~Freire}
\affiliation{Max-Planck-Institut f\"{u}r Radioastronomie, Auf dem H\"{u}gel 69, D-53121 Bonn, Germany}
\author{H.~T.~Cromartie}
\affiliation{University of Virginia, Charlottesville, VA 22904, USA}
\author{S.~M.~Ransom}
\affiliation{National Radio Astronomy Observatory, Charlottesville, VA 22903, USA}
\author{E.~Ferrara}
\affiliation{Goddard Space Flight Center, Greenbelt, MD 20771, USA}
\author{M.~Kerr}
\affiliation{Naval Research Laboratory, Washington, DC 20375, USA}
\author{T.~H.~Burnett}
\affiliation{University of Washington, Seattle, WA 98195, USA}
\author{P.~M.~Saz~Parkinson}
\affiliation{Department of Physics and Laboratory for Space Research, The University of Hong Kong, Pokfulam Road, Hong Kong}
\affiliation{Santa Cruz Institute for Particle Physics, University of California, Santa Cruz, CA, 95064, USA}
%\email{DISTRIBUTION A: Approved for public release, distribution is unlimited.}

\begin{abstract}
We present timing solutions for eight binary millisecond pulsars (MSPs) discovered by searching unidentified \emph{Fermi}-LAT source positions with the 327 MHz receiver of the Arecibo 305-m radio telescope.  { Five of the pulsars are ``spiders'' with orbital periods shorter than 8.1 h. Three of these are in ``black widow'' systems (with degenerate companions of $0.02-0.03~\msun$), one is in a  ``redback'' system (with a non-degenerate companion of $\gtrsim 0.3~\msun$), and one (J1908+2105) is an apparent middle-ground case between the two observational classes.} The remaining three pulsars have white dwarf companions and longer orbital periods. With the initially derived radio timing solutions, we detected $\gamma$-ray pulsations from all MSPs and extended the timing solutions using photons from the full \emph{Fermi} mission, thus confirming the identification of these MSPs with the \emph{Fermi}-LAT sources. The radio emission of the redback is eclipsed during $50\%$ of its orbital period, which is typical for this kind of system. { Two of the black widows exhibit radio eclipses lasting for $10-20\%$ of the orbit, while J1908+2105 eclipses for 40\% of the orbit.
We investigate an apparent link between gamma-ray emission and a short orbital period among known binary MSPs in the Galactic disk, and conclude that selection effects cannot be ruled out as the cause. Based on this analysis we outline how the likelihood of new MSP discoveries can be improved in ongoing and future pulsar searches.} 

\end{abstract}

\section{Introduction}

The \emph{Fermi}-LAT Fourth Source Catalog (4FGL, \citealt{4PC}) contains more than 1300 gamma-ray sources not associated with an object type or a counterpart in another energy band. This constitutes $\sim 26\%$ of all 4FGL sources, which are detected based on eight years of data from the \emph{Fermi} Large Area Telescope (LAT, \citealt{Atwood09}). Since 2008, the first year of the \emph{Fermi} mission, more than 250 pulsars have been detected in gamma rays { \citep{2PC}}\footnote{{ For an up-to-date list, see} \url{https://confluence.slac.stanford.edu/display/GLAMCOG/Public+List+of+LAT-Detected+Gamma-Ray+Pulsars}}, up from $<10$ before \emph{Fermi} was launched. { Many of the pulsars newly detected in gamma rays by \emph{Fermi}-LAT were discovered in blind radio surveys both before and after \emph{Fermi} was launched. However, more than half of the} newly identified gamma-ray pulsars were found in targeted searches of previously unassociated \emph{Fermi}-LAT sources. These are broadly divided into two different classes. The first class consists mostly of young pulsars, which tend to be isolated and have rotation periods of tens to hundreds of milliseconds. In this case it is possible to perform a blind search for gamma-ray pulsations using only \emph{Fermi}-LAT photon data \citep{Abdo09}. This method has produced more than 60 gamma-ray pulsar discoveries to date { (e.g. \citealt{Parkinson10}, \citealt{Pletsch12a})}.

In contrast, millisecond pulsars (MSPs) tend to be in binary systems and have rotation periods of only a few ms.
The short period in combination with the unknown orbital parameters make a blind gamma-ray pulsation search extremely computationally intensive and difficult within a reasonable computing time.
{ Nevertheless, the discovery that known MSPs are an important class of gamma-ray emitters \citep{Abdo09b} has given a crucial impetus to this difficult search.}
In some cases the search can become feasible when { astrometric and} orbital parameters are constrained through optical observations (\citealt{Pletsch12b}, \citealt{Nieder20}, \citealt{Clark20}). { An alternative method relies on searches for radio pulsars at the positions of unidentified {\em Fermi} sources. These surveys have} found more than 80 MSPs ({ e.g.} \citealt{Ransom11}; { \citealt{Cognard11,Keith11}}; \citealt{Ray12,Camilo15,Cromartie16}){ ; these represent about} $25\%$ of all known  MSPs to date in the Galactic disk (hereafter referred to as 'field MSPs' to distinguish them from MSPs residing in globular clusters away from the Galactic plane which are subject to different evolution scenarios and selection biases). After a timing solution is obtained based on 1--2 years of regular radio observations, gamma-ray pulsations are most often detected after \emph{Fermi}-LAT photon arrival times are binned according to the radio ephemeris. If gamma-ray pulsations are detected, in most cases it is possible to extend the timing solution to $>10$ years by using gamma-ray photons and bootstrapping from the shorter span of the radio ephemeris. 

In this paper we present timing solutions and gamma-ray pulsation detections for eight binary MSPs discovered in radio searches of \emph{Fermi}-LAT unassociated sources. Seven of the solutions span the full \emph{Fermi} mission. In one case even though gamma-ray pulsations were detected, it was not possible to extend the timing solution outside the span of radio observations due to rapid quasi-periodic variations in the orbital period.

Section~\ref{sec:radio} describes our radio observations and the data analysis that produces the initial phase-connected timing solutions. Section~\ref{sec:fermi} describes the analysis of gamma-ray photon data and how we extend the timing solutions to the full length of the \emph{Fermi} mission. Section~\ref{sec:results} presents the timing solutions for the pulsars; and Section~\ref{sec:discussion} focuses on the different types of MSP binaries represented in this work and their properties and implications.

\section{Radio Observations and Data Analysis}\label{sec:radio}

The MSPs in this paper were found in targeted searches of unassociated \emph{Fermi}-LAT sources with pulsar-like gamma-ray spectra. \cite{Cromartie16} announced the discovery of six of the MSPs (J0251+2606, J1805+0615, J1824+1014, J1908+2105, J2052+1219, and J1048+2339), and later searches identified two more (J1625$-$0021 and J2006+0148). The search observations used the 327~MHz receiver at the Arecibo telescope. Our subsequent regular timing observations continued with the same setup. We used the Puerto Rican Ultimate Pulsar Processing Instrument (PUPPI), a clone of the GUPPI intrument operating at the Green bank telescope \citep{DuPlain08} in incoherent search mode with a bandwidth of 100~MHz and sampling time of 81.92~$\mu$s. Because the backend bandwidth exceeds the receiver bandwidth, samples were recorded to disk only from the 68.75~MHz within the PUPPI band corresponding to the bandwidth of the 327~MHz receiver. 

During the first month of timing observations, we dedispersed each data set with the dispersion measure (DM) of the pulsar determined during the discovery to remove the frequency-dependent time delay of the pulse across the bandwidth. Then we summed all channels to produce a one-dimensional time series, and used {\tt accelsearch} from PRESTO\footnote{\tt https://www.cv.nrao.edu/\~{}sransom/presto} to perform a Fast Fourier Transform search for periodic signals. This step produces a measurement of the barycentric rotational period ($P_{\rm bary}$) and apparent period derivative, which is related to the line-of-sight acceleration during the observation. 

\subsection{Orbital Solutions}

After the accumulation of several closely spaced observations over the span of a month, it was evident that the MSPs are in binary systems as we observed $P_{\rm bary}$ to vary. This variation is due to Doppler shift induced by orbital motion. We used two methods in succession to obtain an initial fit for the orbital parameters. 
The first method is described by \cite{Freire01} and exploits the fact that for a nearly circular orbit (the most common scenario for MSPs), measurement points in the 2D space defined by $P_{\rm bary}$ and the line-of-sight acceleration fall on an ellipse. This method yields a rough fit to the orbital parameters even with only a handful of observations, as long as they are spread over at least half the orbit. However, the ellipse fit needs measurements of line-of-sight acceleration within individual observations, which are typically only a few minutes long. For one of our MSPs, PSR~J1824+1014, no acceleration was detectable in any of the observations, indicating that the orbital period is on the order of weeks to months. 
In this case, we used a Lomb-Scargle periodogram on the measurements of $P_{\rm bary}$ vs.~time to obtain an initial estimate of the orbital period. Next we used these initial fits as a starting point for fitting a sinusoid to $P_{\rm bary}$ vs.~time in order to refine our estimate of the orbital parameters. 

\subsection{Pulse Times-of-Arrival}

We average (``fold'') the data from each observation with the initial ephemeris, which includes $P_{\rm bary}$, the orbital parameters, and the nominal sky position of the unidentified \emph{Fermi}-LAT source, which is also the center of the discovery radio beam. While individual pulses may differ in shape and height, the average pulse profile is stable in time if areas with extra dispersive delays near eclipse ingress and/or egress are excluded. We fit a set of Gaussians to the folded pulse profile with the highest signal-to-noise ratio. The result is a noiseless multi-Gaussian pulse template, which is then convolved with the average pulse profiles from non-overlapping stretches of data from each observation in order to extract pulse times-of-arrival (TOAs). For each observation, we initially extract 2--5 TOAs from each of two frequency subbands, with each pair of TOAs calculated from 5--10~minutes of data. The specific number of TOAs from each observation is determined by the observation length, the brightness of the detection, and whether the observation includes eclipse ingress or egress, which shortens the usable portion of data.

\subsection{Phase-Connected Timing Solutions}

For each pulsar, we use the TOAs in TEMPO\footnote{\tt http://tempo.sourceforge.net} with the DE421 Solar System ephemeris to obtain a least-squares fit to rotational, astrometric, and orbital parameters, as well as the DM by minimizing the sum of the squares of the differences between actual and expected TOAs. In order to do this we need to determine the correct number of pulsar rotations between TOAs; this is generally achieved using the manual method described in detail by \cite{fr2018}, section 3. However, five of the pulsars (PSRs J1805+0615, J1824+1014, J1908+2105, J2006+0148 and J2052+1219) are only visible when a part of the Galactic plane is visible at Arecibo. The telescope's high over-subscription rate at those times implies that not much time is available to follow up these pulsars, which
results in a relatively sparse data set. In order to determine the correct rotation count between TOAs, we used the ``Dracula'' script, which is described by \cite{fr2018}, section 4. Once the correct rotation counts have been achieved, precise measurements of
the rotational, astrometric, and orbital parameters can be made. At this point, we freeze the DM in the timing solution. We redo the TOA extraction, obtaining only one TOA per observation. Since these TOAs are calculated over the entire frequency band and a longer time span, they have smaller error bars than the TOAs in the initial set and allow us to refine the ephemeris by using them to refit the timing parameters (except DM) with TEMPO. The residuals of this final set of radio TOAs with respect to the post-fit radio timing solution are shown in Figures~\ref{fig_res}--\ref{fig_resorb}.

The radio timing solutions for our eight pulsars have time spans of $2.4-3.4$ years. However, since all pulsars were found in \emph{Fermi}-LAT unassociated sources, we can use our radio timing solutions as a starting point to search for gamma-ray pulsations and extend the timing solution span to the entire $>10$ year span of the \emph{Fermi} mission. 

%\fixme{When redoing par file fits, use EFAC to get a chi-squared near 1.0. --> not necessary if getting Fermi timing solutions}

\subsection{Flux Density Estimates}

\begin{figure}
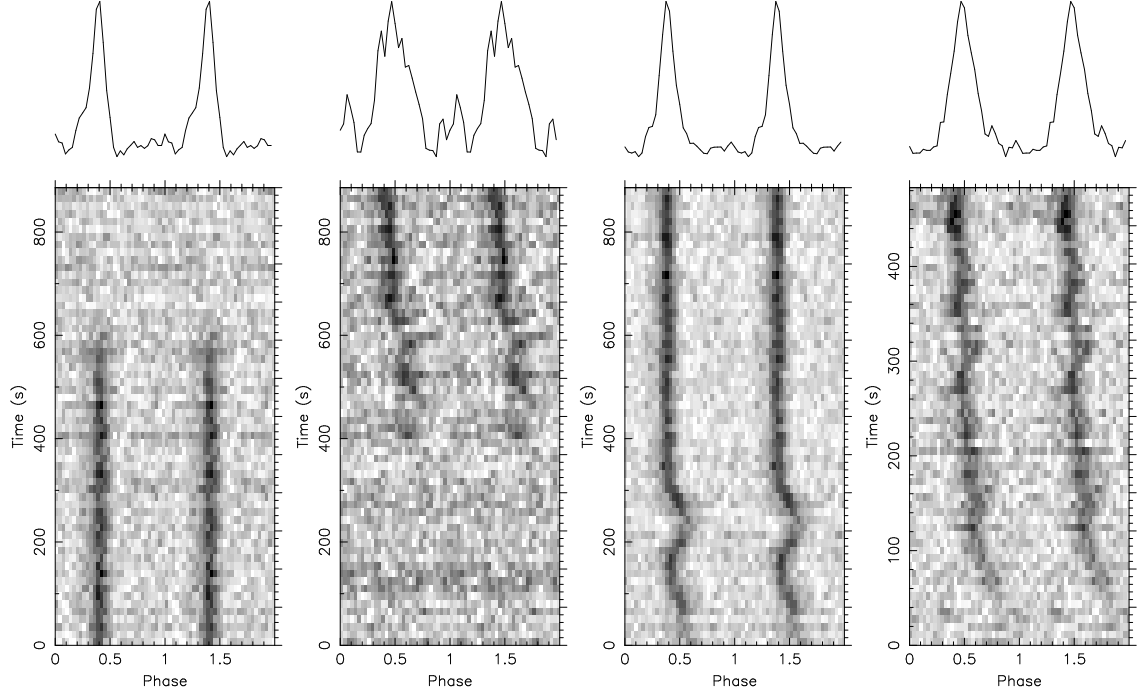

\begin{center}
\includegraphics[angle=-90,width=0.24\textwidth]{fig5-left-to-right-1.eps}
\includegraphics[angle=-90,width=0.24\textwidth]{fig5-left-to-right-2.eps}
\includegraphics[angle=-90,width=0.24\textwidth]{fig5-left-to-right-3.eps}
\includegraphics[angle=-90,width=0.24\textwidth]{fig5-left-to-right-4.eps}
  \caption{Four observations of J2052+1218 made with Arecibo at 327~MHz. The left two panels show ingress and egress on the same orbit (MJD 56650). The respective orbital phase ranges are 0.14--0.23 and 0.32--0.41. The right two panels show egress from two separate observations (MJD 56859 and 57109, orbital phases 0.29--0.38 and 0.36--0.41, respectively). TOAs from ingress and egress regions were not included in the timing solution due to the extra pulse delays. \label{fig_2052_ecl}}
\end{center} 
\end{figure}

Along with the timing solution for each pulsar we give an estimate of the flux density at 327~MHz. For each observation we calculate the uncalibrated period-averaged flux density based on the radiometer equation, the pulse profile template used to extract the TOAs, and using nominal values for the 327~MHz receiver system temperature ($T_{\rm sys}$ = 113~K) and gain ($G = 11$~K/Jy). The recorded value is the average across all observations, excluding non-detections and observations taken near eclipse where the pulsar is not detected during the full integration time (e.g. Figure~\ref{fig_2052_ecl}).

\section{Fermi-LAT Data Analysis}\label{sec:fermi}

We searched for gamma-ray pulsations from each pulsar using Pass 8 \emph{Fermi}-LAT data (\citealt{Atwood13}, \citealt{Bruel18}) from 2008 August 4 to 2019 July 31. We extracted events identified as ``Source'' class ({\tt evclass = 128, evtype = 3}) and detected during periods flagged as producing good science data ({\tt DATA\_QUAL = 1, LAT\_CONFIG = 1}). Additional event cuts were based on zenith angle ($< 90\degree$) and energy (0.1--100~GeV). We extracted events satisfying these criteria  within 15\degree\ of the pulsar's timing position. Then we performed a binned likelihood analysis of a $20\degree \times 20\degree$ region centered on the same position. The sky is modeled using the 4FGL catalog and its corresponding background {\tt gll\_iem\_v07} and {\tt iso\_P8R3\_SOURCE\_V2\_v1.txt}. We fit the spectrum normalization parameters for the two background components. 

If the spectrum of the \emph{Fermi}-LAT source corresponding to the pulsar was originally modeled with a power law, we modified it to a power law with an exponential cut-off, 
\be
\frac{dN}{dE} = N_0 \frac{E^{-\Gamma}}{E_0} ~e^{-E/E_c},
\ee
where $\Gamma$ is the power law index, $E_c$ is the cut-off energy, and $N_0/E_0$ is a normalization constant. For our maximum likelihood fit we used the {\tt P8R3\_SOURCE\_V2} instrument response functions and Science Tools version {\tt 11-00-07}. We fit the spectral parameters of the pulsar source and the normalization parameter of variable sources within 6\degree\ of it. All other parameters were kept fixed at their 4FGL catalog values. Next we used the obtained spectral model to assign weights to photons within 2\degree\ of the pulsar according to the procedure described by \cite{Kerr11}. 

We calculated the photon phases based on the radio timing solution using PINT.\footnote{\tt https://github.com/nanograv/PINT}
%the Fermi plugin for Tempo2\footnote{\tt https://bitbucket.org/psrsoft/tempo2} \citep{Ray11}. \fixme{I actually used fermiphase to assign phases, does it use the same code? --> No, it uses PINT.} 
Since the time span of \emph{Fermi}-LAT data for all our pulsars except J1048+2339 is $> 10$ years, much longer than the $2-3$ year span of the radio timing solutions, the gamma-ray pulse profile drifts in phase outside the range of the radio ephemerides, where the radio timing parameters do not optimally predict the pulsar rotation. We used {\tt event\_optimize} from PINT \citep{Luo19} to obtain 11-year timing solutions using the radio ephemerides as a starting point and applying a Markov Chain Monte Carlo technique for fitting timing model parameters \citep{MCMC}. Based on the H-test significance values of the gamma-ray pulsations, we find the best results when optimizing the position, rotation period, period derivative, and orbital parameters including the orbital period, projected semi-major axis, and time of passage through periastron or ascending node. MCMC-optimized parameters in Tables~\ref{tab:wd}--\ref{tab:rb} are reported as maximum likelihood values with upper and lower uncertainties corresponding to the bounds of the 68\% confidence interval for each parameter. 

The MCMC-optimized timing positions of all eight MSPs are within the 95\% confidence contours of their respective 4FGL sources. All positions except that of J1625$-$0021 are also within the 4FGL 68\% confidence contours. 

\movetabledown=4cm
\begin{table*}
\begin{rotatetable}
\begin{scriptsize}
\caption{Parameters for three pulsar binary systems with likely helium white dwarf companions. All ephemerides use Barycentric Dynamical Time (TDB) units. Uncertainties in parentheses are $1\sigma$ errors from TEMPO fitting. Parameters reported with $\pm$ uncertainties are maximum likelihood values from MCMC optimization. In those cases the uncertainties correspond to the 68\% confidence interval bounds. The pulsar distance is estimated based on the DM and the YMW16 model of Galactic ionized gas density \citep{YMW16}. \label{tab:wd}}
\begin{center}
\begin{tabular}{lccc}
\hline\hline
\multicolumn{4}{c}{Observation and data-set parameters} \\
\hline
Pulsar name\dotfill & J1625$-$0021 & J1824+1014 & J2006+0148 \\ 
MJD range\dotfill & 54682.7--58696.0 & 54682.7--58696.0 & 54682.7--58696.0 \\ 
Data span (yr)\dotfill & 11.0 & 11.0 & 11.0 \\ 
%Number of TOAs\dotfill & 40 &  38  &  32  \\
%Rms timing residual ($\mu s$)\dotfill & 5.8 & 12.2  & 6.6 \\
Reference epoch (MJD)\dotfill & 57114 & 56925 & 57243 \\ 
%Weighted fit\dotfill &  Y \\ 
%Reduced $\chi^2$ value \dotfill & 1.1 \\
\hline
\multicolumn{4}{c}{Measured Quantities} \\ 
\hline
Right ascension, $\alpha$ (hh:mm:ss)\dotfill & 16:25:10.3579$^{+0.0002}_{-0.0003}$ & 18:24:14.933$^{+0.001}_{-0.003}$ & 20:06:29.0529$^{+0.0002}_{-0.0001}$ \\ 
Declination, $\delta$ (dd:mm:ss)\dotfill & $-$00:21:28.960$^{+0.008}_{-0.008}$ & +10:14:43.82$^{+0.02}_{-0.03}$ & +01:48:53.919$^{+0.007}_{-0.006}$ \\ 
Dispersion measure, DM (cm$^{-3}$pc)\dotfill & 16.497 & 59.8807 &  43.5786 \\ 
Pulse frequency, $\nu$ (s$^{-1}$)\dotfill & 352.90623328676$^{+2\times 10^{-11}}_{-3\times 10^{-11}}$ &  245.9470303117$^{+3\times10^{-10}}_{-1\times10^{-10}}$ & 462.20187192375$^{+3\times10^{-11}}_{-3\times10^{-11}}$ \\ 
First derivative of pulse frequency, $\dot{\nu}$ ($10^{-15}\, \rm Hz \, s^{-1}$)\dotfill & $-$2.6559$^{+0.0004}_{-0.0005}$ & $-$0.3247$^{+0.0009}_{-0.003}$ & $-$0.703$^{+0.008}_{-0.003}$ \\ 
Proper motion in right ascension, $\mu_{\alpha} \cos \delta$ (mas\,yr$^{-1}$)\dotfill & $-$3.6(13) & - & - \\ 
Proper motion in declination, $\mu_{\delta}$ (mas\,yr$^{-1}$)\dotfill & $-$8.9(42) & - & - \\ 
Binary model\dotfill & ELL1 & DD & ELL1 \\
Orbital period, $P_{\rm b}$ (d)\dotfill & 7.38297056$^{+1\times10^{-8}}_{-1\times10^{-8}}$ &  82.548980$^{+2\times10^{-6}}_{-1\times10^{-6}}$ &  0.6513597615$^{+4\times10^{-10}}_{-6\times10^{-10}}$ \\ 
Projected semi-major axis of pulsar's orbit, $x$ (lt-s)\dotfill & 4.584723$^{+6\times10^{-6}}_{-8\times10^{-6}}$ &  35.44585$^{+3\times10^{-5}}_{-4\times10^{-5}}$ & 0.842670$^{+4\times10^{-6}}_{-8\times10^{-6}}$  \\ 
Orbital eccentricity, $e$ \dotfill & - & 1.376(2)$\times 10^{-4}$ & - \\ 
Time of passage through periastron, $T_0$ (MJD)\dotfill & - &  56523.10321$^{+1\times10^{-5}}_{-6\times10^{-5}}$ & - \\ 
Longitude of periastron, $\omega$ ($\deg$) \dotfill & - &  301.06(8) & - \\ 
Time of ascending node, $T_{\rm asc}$ (MJD)\dotfill &  57154.884218$^{+2\times10^{-6}}_{-2\times10^{-6}}$ & - &  57243.4389465$^{+3\times10^{-7}}_{-1\times10^{-6}}$ \\ 
EPS1\dotfill & $-$2.56(87)$\times 10^{-6}$ & - & 2.03(76)$\times 10^{-5}$ \\ 
EPS2\dotfill & $-$0.64(74)$\times 10^{-6}$ & - & 6.1(38)$\times 10^{-6}$  \\
\hline
\multicolumn{4}{c}{Derived Quantities} \\
\hline
Galactic longitude, $l$ ($\deg$) \dotfill & 13.894 & 39.076 & 43.390 \\
Galactic latitude, $b$ ($\deg$) \dotfill & 31.831  & 10.647 & -15.765 \\
Distance, $D$ (kpc) \dotfill & 0.95 & 2.9 & 2.4 \\
Spin period, $P$ (s) \dotfill & 0.00283361387722 & 0.00406591613947 & 0.00216355679357 \\ 
Spin period derivative, $\dot{P}$ ($10^{-21}\, \rm s \, s^{-1}$) \dotfill & 21.33 & 5.37 & 3.29 \\   
$\log_{10}$(Characteristic age, yr) \dotfill & 9.32 & 10.08 & 10.02 \\
$\log_{10}$(Surface magnetic field strength, G) \dotfill & 8.40 & 8.17 & 7.93  \\
$\log_{10}$(Edot, ergs/s) \dotfill & 34.57 & 33.50 & 34.11  \\
Mass function, $f$ ($M_{\odot}$) \dotfill & 0.0019 & 0.0070 & 0.0015 \\
Minimum companion mass, $M_c$ ($M_{\odot}$) \dotfill & 0.17 & 0.27 & 0.15 \\
Uncalibrated flux density at 327~MHz, $S_{\rm 327}$(mJy) \dotfill & 0.23 & 0.13 & 0.25 \\
\hline
%\multicolumn{2}{c}{Assumptions} \\
%\hline
%Clock correction procedure\dotfill & TT(TAI) \\
%Solar system ephemeris model\dotfill & DE421 \\
%Binary model\dotfill & ELL1 \\
%TDB units (tempo1 mode)\dotfill & Y \\
%FB90 time ephemeris (tempo1 mode)\dotfill & Y \\
%T2C (tempo1 mode)\dotfill & Y \\
%Shapiro delay due to planets\dotfill & N \\
%Tropospheric delay\dotfill & N \\
%Dilate frequency\dotfill & N \\
%Electron density at 1 AU (cm$^{-3}$)\dotfill & 10.00 \\ 
%Model version number\dotfill & 2.00 \\ 
%\hline
\end{tabular}
\end{center}
\end{scriptsize}
\end{rotatetable}
\end{table*}

% \movetableright=-3cm
\movetabledown=8cm
\begin{table}
\begin{rotatetable}
\begin{scriptsize}
\begin{center}
\caption{Parameters for the ``black widow" pulsars. Notation as in Table~\ref{tab:wd}. { J1908+2105 spans the observational properties of the black widow and redback classes. Here and in Figure~\ref{fig:pb-mc} it is provisionally grouped with black widows as its minimum companion mass is closer to what is typical for this type of binary system.}\label{tab:bw}}
\begin{tabular}{lcccc}
\hline\hline
\multicolumn{5}{c}{Observations and data-set parameters} \\
\hline
Pulsar name\dotfill & J0251+2606 & J1805+0615 & J1908+2105 & J2052+1219 \\ 
MJD range\dotfill & 54682.7--58696.0 & 54682.7--58696.0 & 54682.7--58696.0 & 54682.7--58696.0 \\ 
Data span (yr)\dotfill & 11.0 & 11.0 & 11.0 & 11.0 \\ 
%Number of TOAs\dotfill & 64 & 39 & 49 \\
%Rms timing residual ($\mu s$)\dotfill & 11.9 & 9.1 & 5.0 \\
Reference epoch (MJD)\dotfill & 56644 & 57002 & 56930 & 56548 \\
%Weighted fit\dotfill &  Y \\ 
%Reduced $\chi^2$ value \dotfill & 1.3 \\
\hline
\multicolumn{5}{c}{Measured Quantities} \\ 
\hline
Right ascension, $\alpha$ (hh:mm:ss)\dotfill & 02:51:02.5540$^{+0.0003}_{-0.0007}$ & 18:05:42.39969$^{+3\times10^{-5}}_{-1\times10^{-3}}$ & 19:08:57.2939$^{+0.0004}_{-0.0002}$ & 20:52:47.77818 $^{+9\times10^{-5}}_{-3\times10^{-4}}$ \\ 
Declination, $\delta$ (dd:mm:ss)\dotfill & +26:06:09.97$^{+0.02}_{-0.02}$ & +06:15:18.614$^{+0.005}_{-0.020}$ & +21:05:02.720$^{+0.008}_{-0.003}$ & +12:19:59.020$^{+0.006}_{-0.003}$ \\ 
Proper motion in right ascension, $\mu_{\alpha} \cos \delta$ (mas\,yr$^{-1}$)\dotfill & 17(3) & 8.7(13) & - & $-$4.30(32) \\ 
Proper motion in declination, $\mu_{\delta}$ (mas\,yr$^{-1}$)\dotfill & - & 12.8(29) & - & $-$13.96(59) \\ 
Dispersion measure, DM (cm$^{-3}$pc)\dotfill & 20.2166 & 64.883 & 61.9067 & 41.9596 \\ 
Pulse frequency, $\nu$ (s$^{-1}$)\dotfill &  393.46001048595$^{+2\times10^{-11}}_{-5\times10^{-11}}$ &  469.72472452336$^{+8\times10^{-11}}_{-1\times10^{-10}}$ & 389.955879788271$^{+6\times10^{-11}}_{-9\times10^{-12}}$ & 503.71330349547$^{+1\times10^{-11}}_{-3\times10^{-11}}$ \\ 
First derivative of pulse frequency, $\dot{\nu}$ ($10^{-15}\, \rm Hz \, s^{-1}$)\dotfill & $-$1.1713$^{+0.002}_{-0.0002}$ & $-$5.0216$^{+0.0005}_{-0.002}$ & $-$2.1038$^{+0.0006}_{-0.0006}$ & $-$1.7009$^{+0.0004}_{-0.0005}$ \\
Binary model\dotfill & ELL1 & ELL1 & ELL1 & ELL1 \\ %& ELL1 \\
Orbital period, $P_{\rm b}$ (d)\dotfill & 0.2024406405$^{+7\times10^{-10}}_{-1\times10^{-9}}$ &  0.3368720352$^{+6\times10^{-10}}_{-9\times10^{-9}}$ & 0.1463168431$^{+6\times10^{-10}}_{-1\times10^{-10}}$ & 0.1146136250$^{+2\times10^{-10}}_{-1\times10^{-10}}$ \\ 
Projected semi-major axis of orbit, $x$ (lt-s)\dotfill & 0.065678$^{+1\times10^{-5}}_{-5\times10^{-6}}$ & 0.087733$^{+9\times10^{-6}}_{-2\times10^{-5}}$ & 0.116895$^{+2\times10^{-5}}_{-2\times10^{-6}}$ & 0.061375$^{+6\times10^{-6}}_{-2\times10^{-6}}$ \\
Time of ascending node, $T_\mathrm{asc}$ (MJD)\dotfill & 56647.008791$^{+5\times10^{-6}}_{-4\times10^{-6}}$ & 57001.602777$^{+9\times10^{-6}}_{-4\times10^{-6}}$ & 56478.283490$^{+2\times10^{-6}}_{-2\times10^{-6}}$ & 56548.106219$^{+1\times10^{-6}}_{-2\times10^{-6}}$ \\
EPS1\dotfill & $-$3.1(76)$\times 10^{-5}$ & 1.55(43)$\times 10^{-4}$ & 0 & 0 \\ 
EPS2\dotfill & $-$1.17(65)$\times 10^{-4}$ & 1.0(36)$\times 10^{-5}$ & 0 & 0 \\ 
\hline
\multicolumn{5}{c}{Derived Quantities} \\
\hline
Galactic longitude, $l$ ($\deg$) \dotfill & 153.881 & 33.351 & 53.690 & 59.139 \\
Galactic latitude, $b$ ($\deg$) \dotfill & $-$29.495 & 13.006 & 5.772 & $-$19.989 \\
Distance, $D$ (kpc) \dotfill & 0.96 & 3.9 & 2.6 & 3.9 \\ 
Spin period, $P$ (s) \dotfill & 0.00254155434695 & 0.00212890645902 & 0.00256439267063 & 0.00198525628182 \\ 
Spin period derivative, $\dot{P}$ ($10^{-21}\, \rm s \, s^{-1}$) \dotfill & 7.57 & 22.76 & 13.84 & 6.70 \\ 
$\log_{10}$(Characteristic age, yr) \dotfill & 9.73 & 9.17 & 9.47 & 9.67 \\ 
$\log_{10}$(Surface magnetic field strength, G) \dotfill & 8.14 & 8.35 & 8.28 & 8.07 \\
$\log_{10}$(Edot, ergs/s) \dotfill & 34.26 & 34.97 & 34.51 & 34.53 \\ 
Mass function, $f$ ($M_{\odot}$) \dotfill & $7.4\times10^{-6}$ & $6.4\times10^{-6}$ & $8.0\times10^{-5}$ & $1.9\times10^{-5}$  \\ 
Minimum companion mass, $M_c$ ($M_{\odot}$) \dotfill & 0.025 & 0.023 & 0.055 & 0.034 \\ 
Uncalibrated flux density at 327~MHz, $S_{\rm 327}$(mJy) \dotfill & 0.17 & 0.43 & 0.19 & 0.52 \\ 
\hline
%\multicolumn{2}{c}{Assumptions} \\
%\hline
%Clock correction procedure\dotfill & TT(TAI) \\
%Solar system ephemeris model\dotfill & DE405 \\
%Binary model\dotfill & ELL1 \\
%TDB units (tempo1 mode)\dotfill & Y \\
%FB90 time ephemeris (tempo1 mode)\dotfill & Y \\
%T2C (tempo1 mode)\dotfill & Y \\
%Shapiro delay due to planets\dotfill & N \\
%Tropospheric delay\dotfill & N \\
%Dilate frequency\dotfill & N \\
%Electron density at 1 AU (cm$^{-3}$)\dotfill & 10.00 \\ 
%Model version number\dotfill & 2.00 \\ 
\hline
\end{tabular}
\end{center}
\end{scriptsize}
\end{rotatetable}
\end{table}

\movetableright=-2cm
\begin{table}
\begin{scriptsize}
\begin{center}
\caption{Parameters for the eclipsing ``redback'' pulsar J1048+2339. Notation as in Table~\ref{tab:wd}. \label{tab:rb}}
\begin{tabular}{lc}
\hline\hline
\multicolumn{2}{c}{Observation and data-set parameters} \\
\hline
Pulsar name\dotfill & J1048+2339 \\ 
MJD range\dotfill & 56508.8---57548.9 \\ 
Data span (yr)\dotfill & 2.85 \\ 
Epoch of frequency determination (MJD)\dotfill & 57285  \\ 
Epoch of position determination (MJD)\dotfill & 56700 \\ 
Epoch of dispersion measure determination (MJD)\dotfill & 56700 \\ 
\hline
\multicolumn{2}{c}{Measured Quantities} \\ 
\hline
Right ascension, $\alpha$ (hh:mm:ss)\dotfill &  10:48:43.418354(76) \\ 
Declination, $\delta$ (dd:mm:ss)\dotfill & +23:39:53.4043(20) \\ 
Proper motion in $\alpha$, $\mu_{\alpha} \cos \delta$ (mas\,yr$^{-1}$)\dotfill & $-$18.7\tablenotemark{a} \\ 
Proper motion in $\delta$, $\mu_{\delta}$ (mas\,yr$^{-1}$)\dotfill & $-$9.4\tablenotemark{a} \\
Dispersion measure, DM (cm$^{-3}$pc)\dotfill & 16.6543 \\ 
Pulse frequency, $\nu$ (s$^{-1}$)\dotfill & 214.35478534113(2) \\ 
First derivative of $\nu$, $\dot{\nu}$ ($10^{-15}\, \rm Hz\, s^{-2}$)\dotfill & $-$1.382(2) \\ 
Binary model\dotfill & BTX \\
Orbital period (d) \dotfill & 0.2505191\tablenotemark{b} \\ 
Projected semi-major axis of orbit, $x$ (lt-s)\dotfill & 0.836120(3) \\ 
Epoch of periastron, $T_0$ (MJD)\dotfill & 56637.598174(2) \\ 
Longitude of periastron, $\omega_0$ (deg)\dotfill & 0 \\ 
Orbital eccentricity, $e$\dotfill & 0 \\ 
Orbital frequency, $FB0$ (Hz)\dotfill & 4.6200355(5)$\times 10^{-5}$ \\ 
1st orbital frequency derivative, $FB1$ (Hz~s$^{-1}$)\dotfill & $-$4.18(55)$\times 10^{-18}$ \\ 
2nd orbital frequency derivative, $FB2$ (Hz~s$^{-2}$)\dotfill & 1.4(12)$\times 10^{-25}$ \\ 
3rd orbital frequency derivative, $FB3$ (Hz~s$^{-3}$)\dotfill & $-$1.50(61)$\times 10^{-31}$ \\ 
4th orbital frequency derivative, $FB4$ (Hz~s$^{-4}$)\dotfill & 8.0(13)$\times 10^{-38}$ \\ 
5th orbital frequency derivative, $FB5$ (Hz~s$^{-5}$)\dotfill & $-$1.65(18)$\times 10^{-44}$ \\ 
6th orbital frequency derivative, $FB6$ (Hz~s$^{-6}$)\dotfill & 1.83(17)$\times 10^{-51}$ \\ 
7th orbital frequency derivative, $FB7$ (Hz~s$^{-7}$)\dotfill & $-$1.119(91)$\times 10^{-58}$ \\ 
8th orbital frequency derivative, $FB8$ (Hz~s$^{-8}$)\dotfill & 3.00(23)$\times 10^{-66}$ \\ 
\hline
\multicolumn{2}{c}{Derived Quantities} \\
\hline
Galactic longitude, $l$ ($\deg$) \dotfill & 32.709 \\
Galactic latitude, $b$ ($\deg$) \dotfill & 62.139 \\
Distance, $D$ (kpc) \dotfill & 1.7 \\
Spin period, $P$ (s) \dotfill & 0.00466516293727 \\ 
Spin period derivative, $\dot{P}$ ($10^{-21}\, \rm s \, s^{-1}$) \dotfill & 30.07 \\   
$\log_{10}$(Characteristic age, yr) \dotfill & 9.39 \\
$\log_{10}$(Surface magnetic field strength, G) \dotfill & 8.58 \\
$\log_{10}$(Edot, ergs/s) \dotfill & 34.07 \\
Mass function, $f$ ($M_{\odot}$) \dotfill & 0.01 \\
Minimum companion mass, $M_c$ ($M_{\odot}$) \dotfill & 0.31 \\
Uncalibrated flux density at 327~MHz, $S_{\rm 327}$(mJy) \dotfill & 0.42 \\
\hline
%\multicolumn{2}{c}{Assumptions} \\
%\hline
%Clock correction procedure\dotfill & TT(TAI) \\
%Solar system ephemeris model\dotfill & DE421 \\
%Binary model\dotfill & BTX \\
%TDB units (tempo1 mode)\dotfill & Y \\
%FB90 time ephemeris (tempo1 mode)\dotfill & Y \\
%T2C (tempo1 mode)\dotfill & Y \\
%Shapiro delay due to planets\dotfill & N \\
%Tropospheric delay\dotfill & N \\
%Dilate frequency\dotfill & N \\
%Electron density at 1 AU (cm$^{-3}$)\dotfill & 10.00 \\ 
%Model version number\dotfill & 2.00 \\ 
\hline
\end{tabular}
\end{center}
\tablenotetext{a}{The proper motion component measurements are from optical observations and are used as fixed parameters in the timing solution (see \citealt{Deneva16}, Section~6.2 for detailed discussion).}
\tablenotetext{b}{Since the orbital period is variable, this is an approximation calculated as 1/$FB0$.}
\end{scriptsize}
\end{table}

\begin{figure}
\begin{center}
  \includegraphics[width=0.99\textwidth]{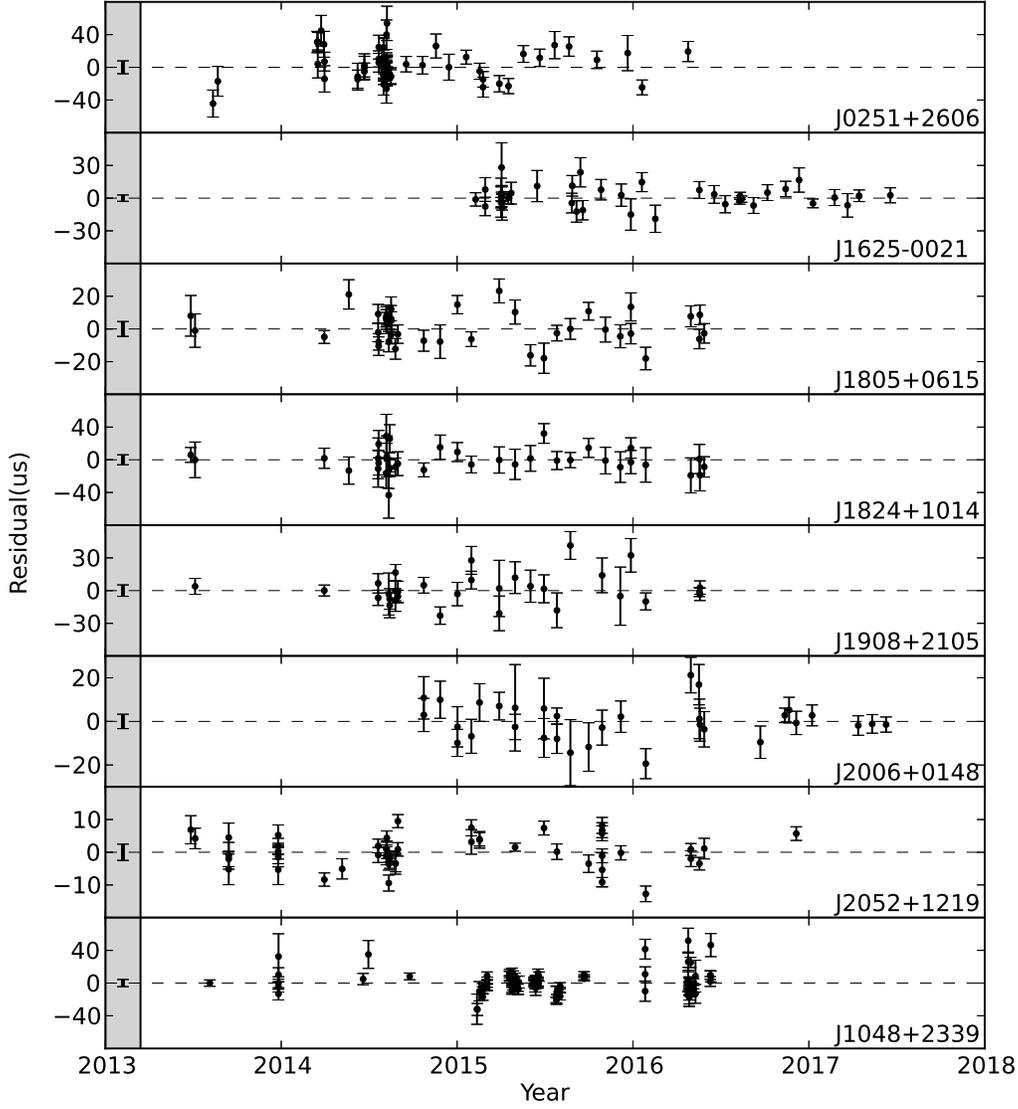}
\caption{Radio timing residuals vs.~observing epoch. All observations are at 327~MHz. The error bar within the shaded area on the left of each panel shows the size of the weighted root-mean-square residual. \label{fig_res}}
\end{center} 
\end{figure}

\begin{figure}
\begin{center}
  \includegraphics[width=0.99\textwidth]{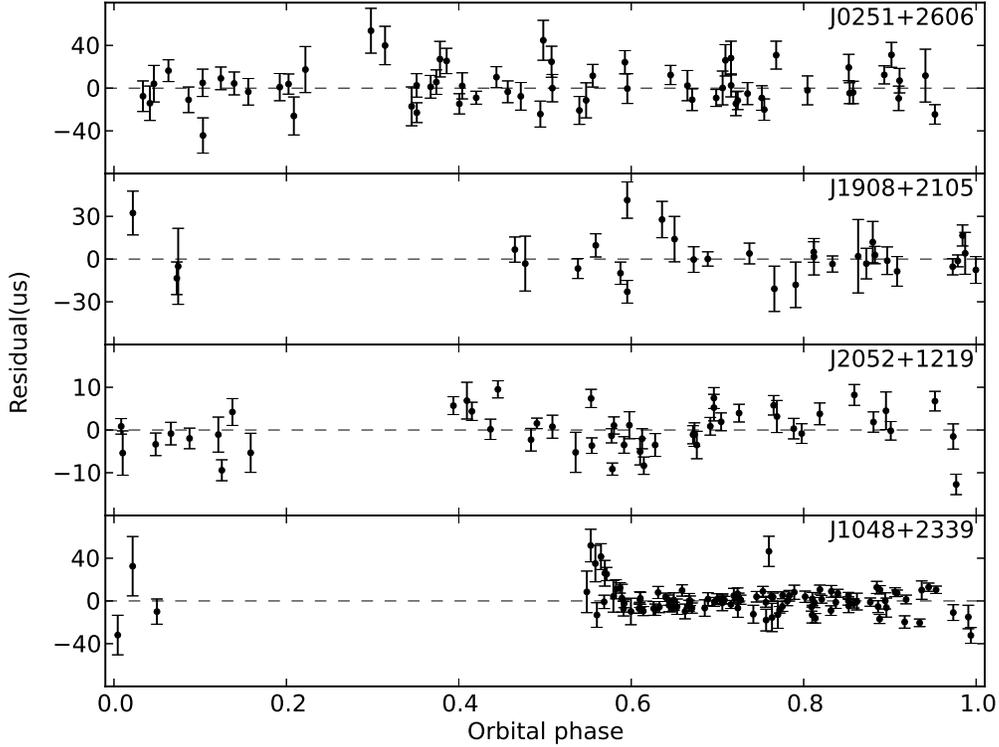}
\caption{Radio timing residuals vs.~orbital phase for the four eclipsing pulsars in this work. All observations are at 327~MHz. \label{fig_resorb}}
\end{center} 
\end{figure}

%\begin{landscape}
%\begin{figure}
%\begin{center}
%\includegraphics[width=0.24\textwidth]{0251-phaseogram.eps}
%\includegraphics[width=0.24\textwidth]{1625-phaseogram.eps}
%\includegraphics[width=0.24\textwidth]{1805-phaseogram.eps}
%\includegraphics[width=0.24\textwidth]{1824-phaseogram.eps}
%\caption{weibpqigeb}
%\end{center}
%\end{figure}
%\end{landscape}

\begin{figure}
\begin{center}
\includegraphics[width=0.49\textwidth]{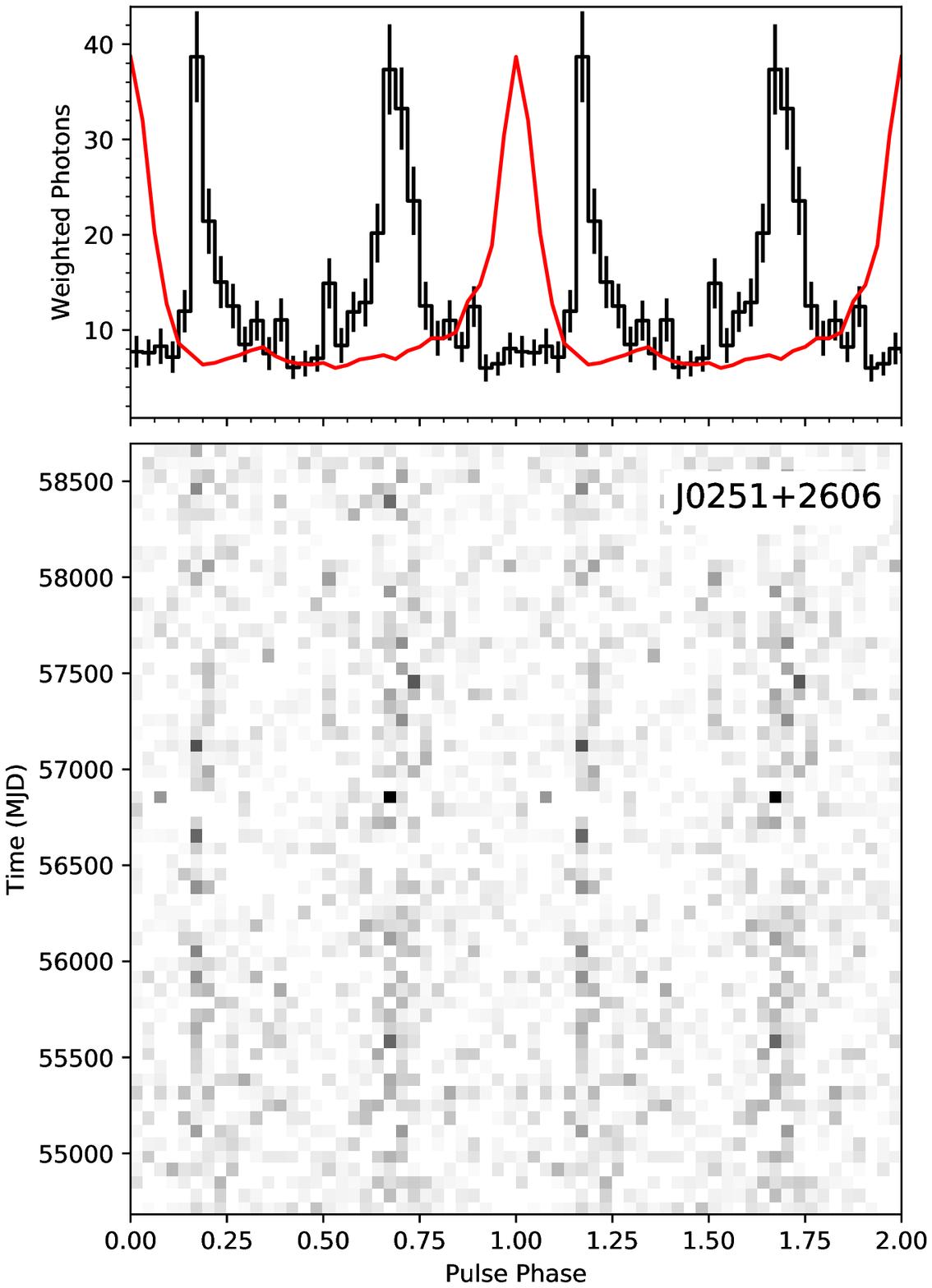}
\includegraphics[width=0.49\textwidth]{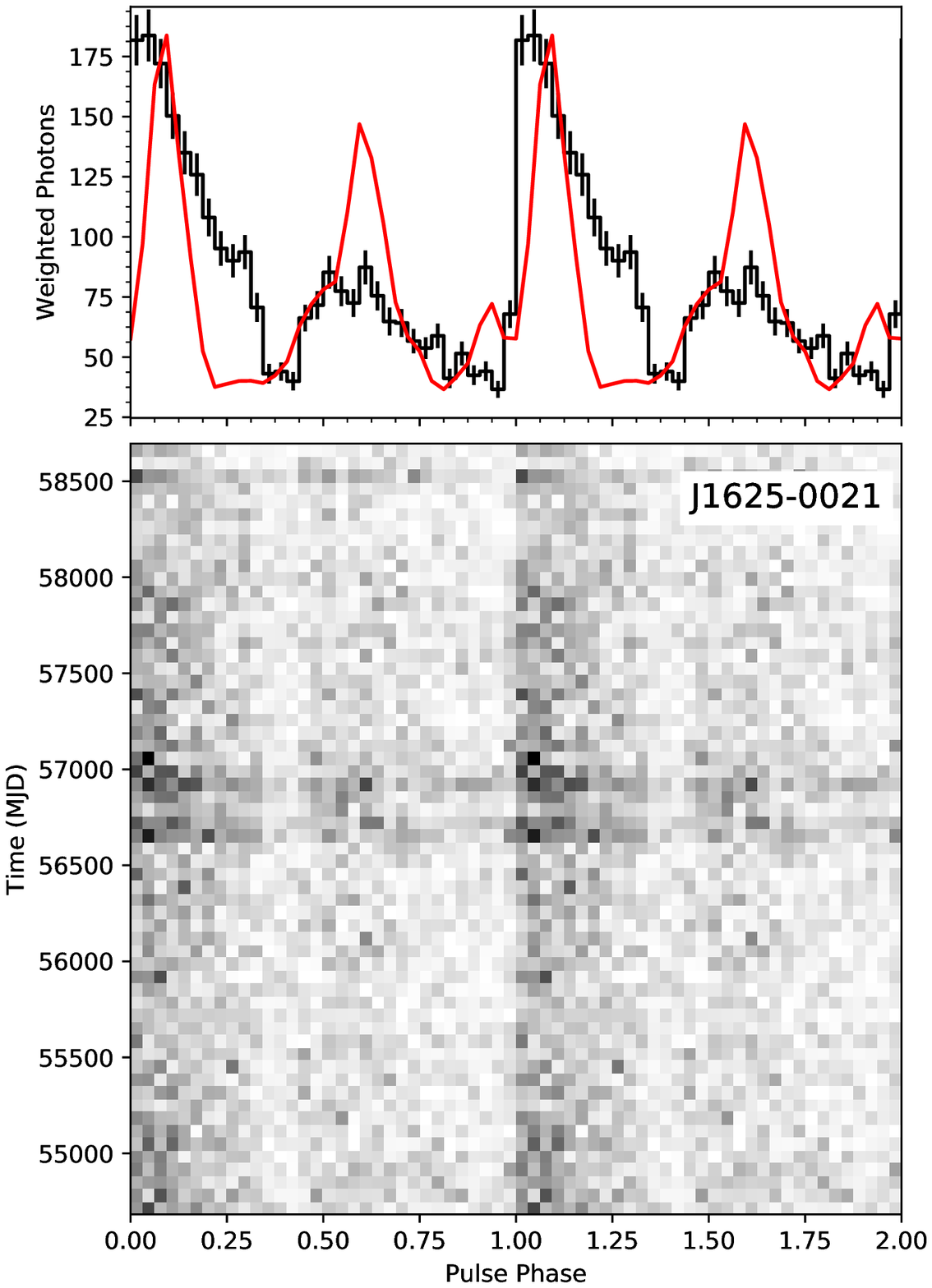}
% \caption{Weighted \emph{Fermi} photon phaseograms (bottom) and average pulse profiles (top) for PSRs~J0251+2606 (left) and J1625-0021 (right). The phase-aligned 327~MHz radio pulse profile is shown in red in the top panel.\label{phaseogram_0251_1625}}
% \end{center}
% \end{figure}

% \begin{figure}
% \begin{center}
%\includegraphics[width=0.49\textwidth]{1805-phaseogram.eps}
%\includegraphics[width=0.49\textwidth]{1824-phaseogram.eps}
\includegraphics[width=0.49\textwidth]{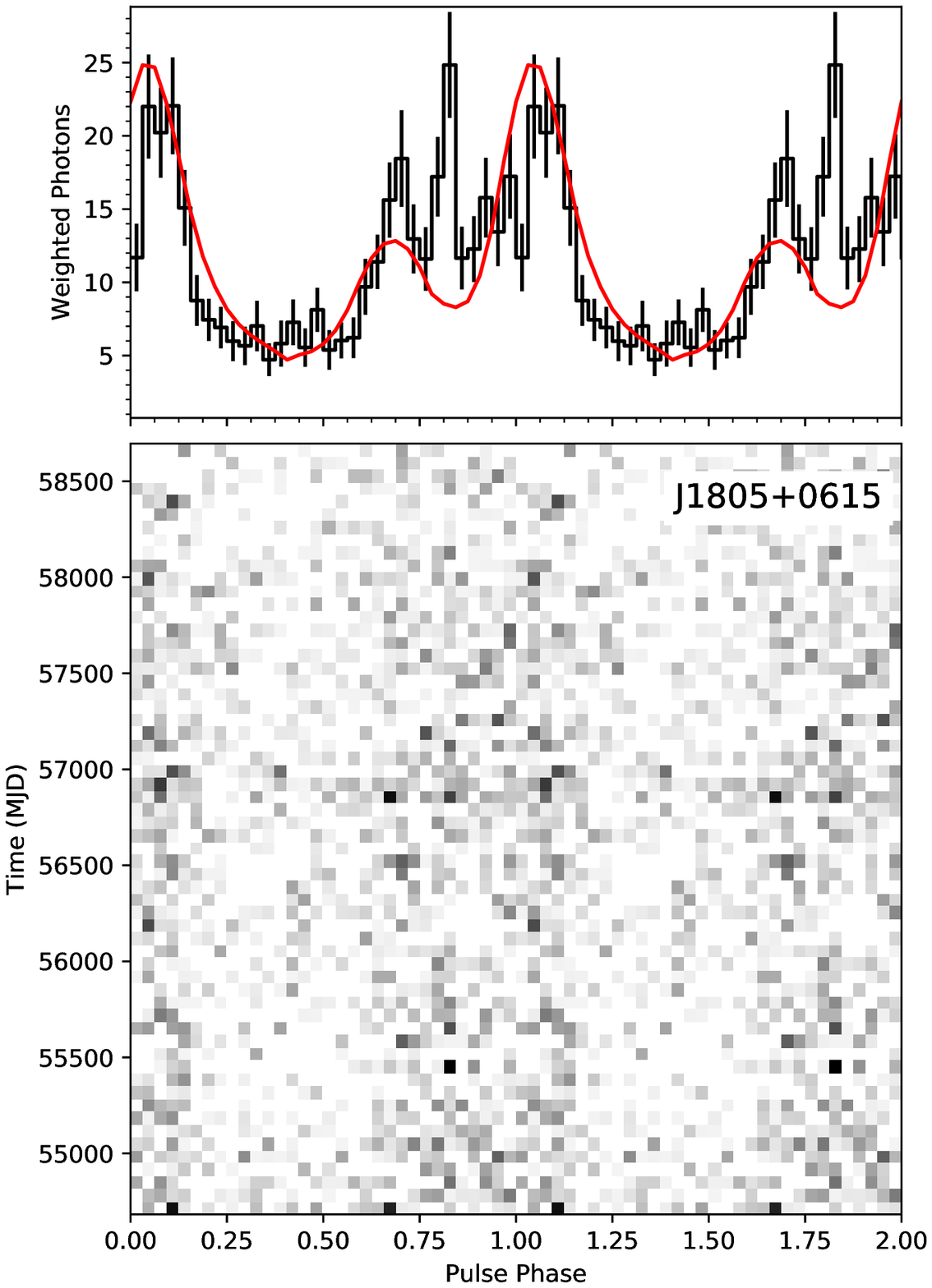}
\includegraphics[width=0.49\textwidth]{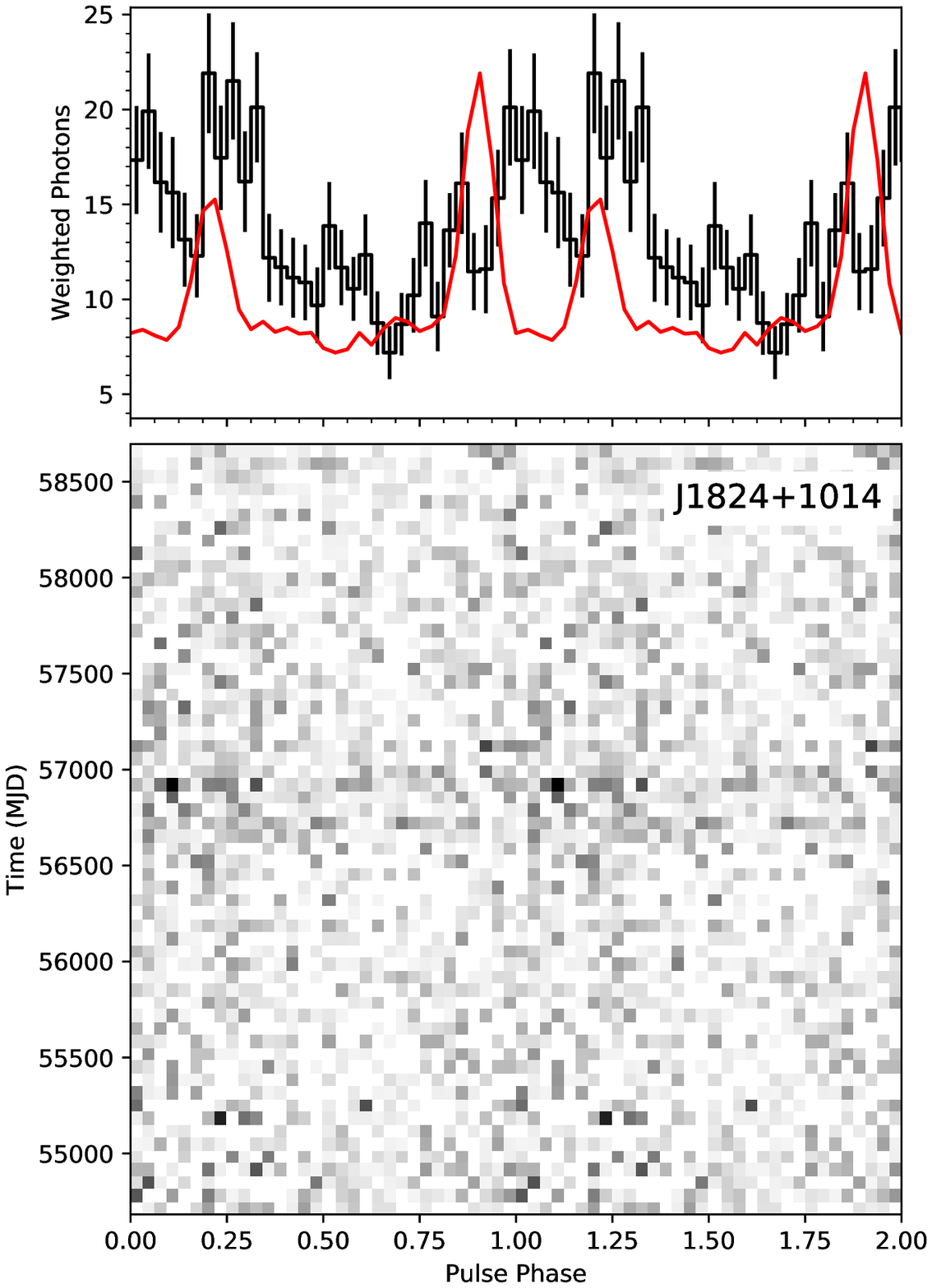}
\caption{Weighted \emph{Fermi}-LAT photon phaseograms (lower panels) and average pulse profiles (top panels) for PSRs~J0251+2606 (upper left), J1625-0021 (upper right), J1805+0615 (lower left) and J1824+1014 (lower right). The phase-aligned 327~MHz radio pulse profile is shown in red in the top panels. A temporary change in the \emph{Fermi} all-sky survey strategy at MJD~56500--57000 caused a change in average exposure depending on source position. This is visible as darker horizontal bands in some of the greyscale plots. \label{phaseogram_1}}
\end{center}
\end{figure}

\begin{figure}
\begin{center}
\includegraphics[width=0.49\textwidth]{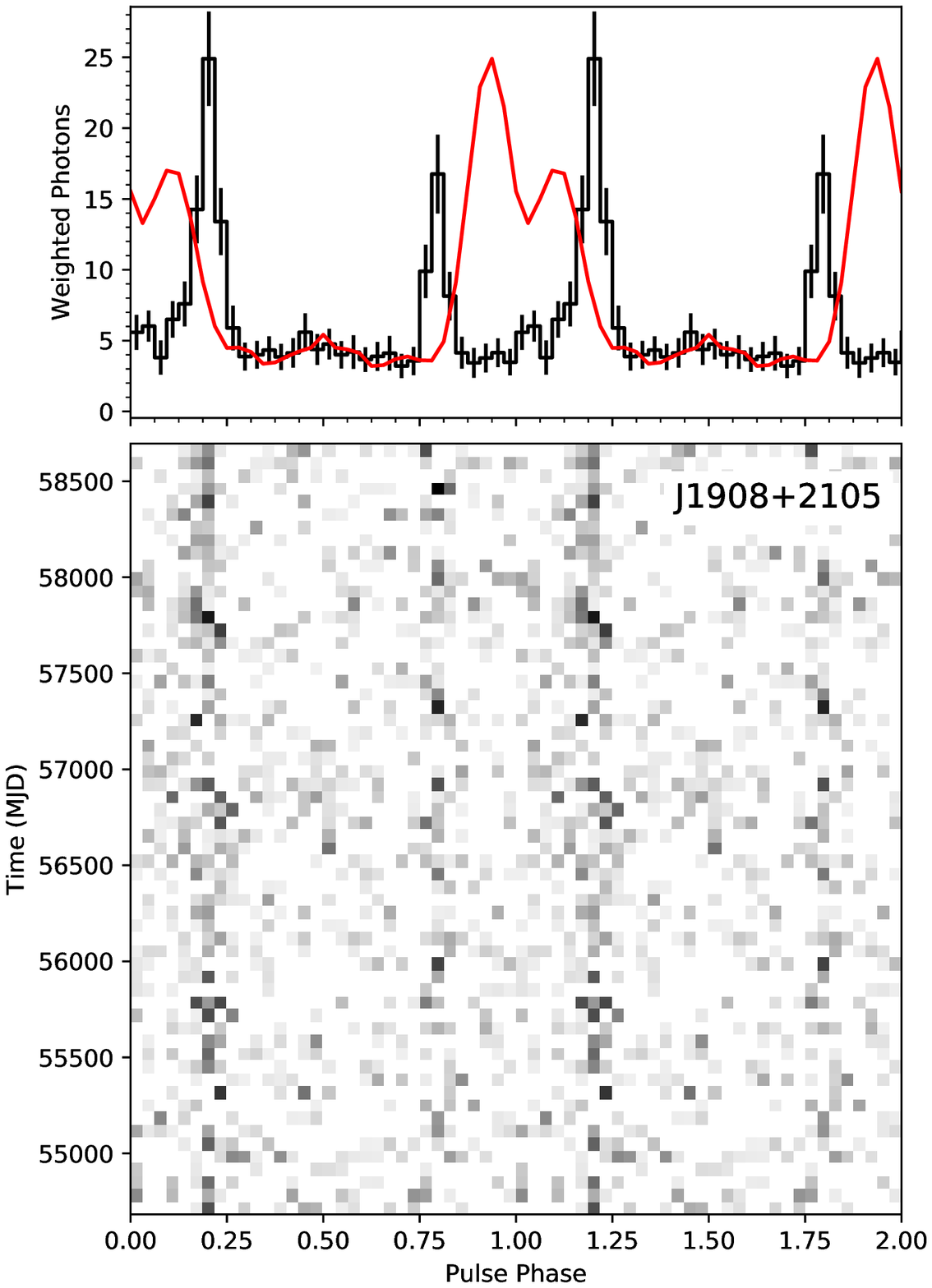}
\includegraphics[width=0.49\textwidth]{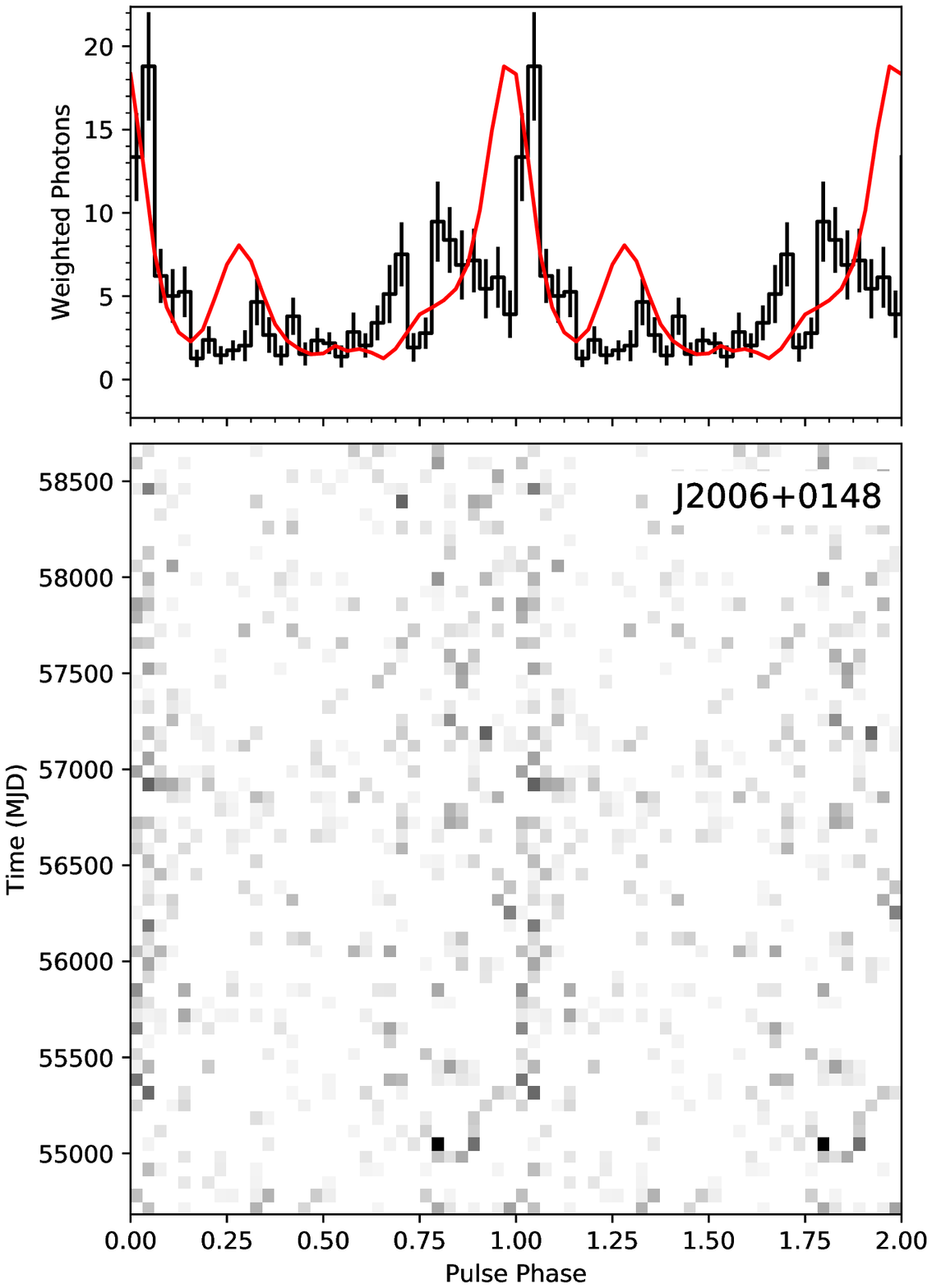}
% \caption{Weighted \emph{Fermi} photon phaseograms (bottom) and average pulse profiles (top) for PSRs~J1908+2105 (left) and ). The phase-aligned 327~MHz radio pulse profile is shown in red in the top panel.\label{phaseogram_1908_2006}}
% \end{center}
% \end{figure}

% \begin{figure}
% \begin{center}
%\includegraphics[width=0.49\textwidth]{2052-phaseogram.eps}
%\includegraphics[width=0.49\textwidth]{1048-phaseogram.eps}
\includegraphics[width=0.49\textwidth]{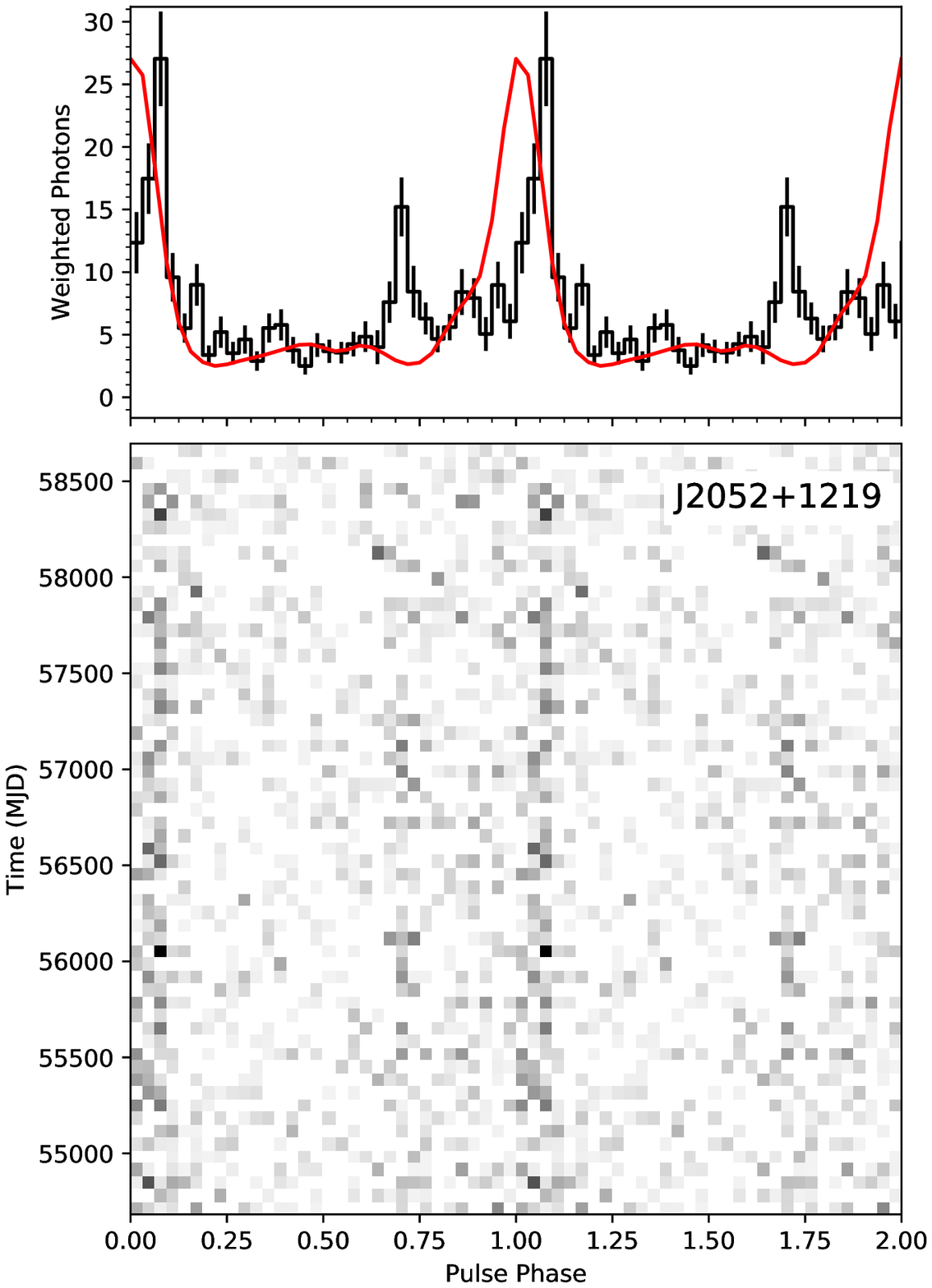}
\includegraphics[width=0.49\textwidth]{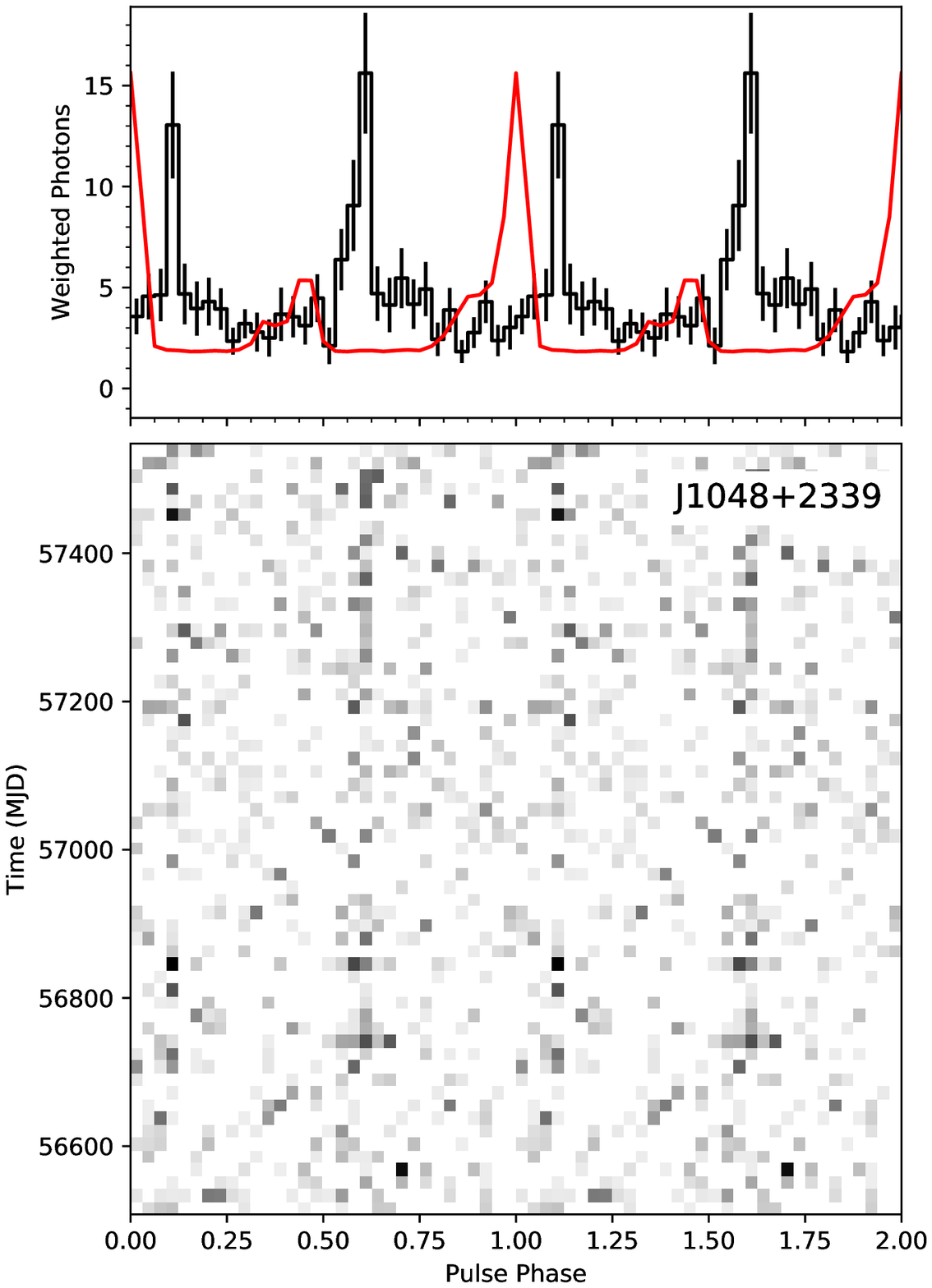}
\caption{Same as Figure~\ref{phaseogram_1}, for PSRs~J1908+2105 (top left), J2006+0148 (top right), J2052+1219 (lower left) and J1048+2339 (lower right).\label{phaseogram_2}}
\end{center}
\end{figure}

\section{Results}\label{sec:results}

%\fixme{Check for X-ray and optical counterparts (except 1048, that was already covered in 1048 paper).} --> Checked on SIMBAD. None have matches within the timing position uncertainty. The closest is for J1625: X-ray source J1625 X1, which is 7.4 arcsec away; but the timing position error is <0.5 arcsec.

\subsection{PSR~J0251$+$2606}

This eclipsing black widow MSP was discovered in the \emph{Fermi}-LAT source 3FGL~J0251.1+2603 (4FGL~J0251.0+2605). The radio timing residuals vs.~observing epoch and orbital phase are shown in Figures~\ref{fig_res} and \ref{fig_resorb}. The \emph{Fermi} timing solution is shown in Table~\ref{tab:bw}. The pulsar has an orbital period of 4.86 hours and a minimum companion mass of 0.025~\msun. Using the $\sim 3$-yr radio timing solution as a starting point and optimizing it based on \emph{Fermi}-LAT photon data from the full 11-year gamma-ray data set, we were able to detect gamma-ray pulsations with a H-test significance of 383.08 (18$\sigma$). Figure~\ref{phaseogram_1} (top left) shows the photon phaseogram and folded gamma-ray profile phase-aligned with the 327~MHz radio profile. The alignment is done using PINT to calculate photon phases with respect to the same phase reference used for extracting radio TOAs and to construct an average gamma-ray pulse profile. 

%Read 36668 events
%htest = 381.42, significance = 17.95 sigma

\subsection{PSR~J1625$-$0021}

This MSP was discovered in the \emph{Fermi}-LAT source 3FGL~J1625.1-0021 (4FGL~J1625.1-0020). The radio timing residuals vs.~observing epoch are shown in Figure~\ref{fig_res}. The \emph{Fermi} timing solution is shown in Table~\ref{tab:wd}. The pulsar has an orbital period of 7.4 days and a minimum companion mass of 0.17~\msun. The high companion mass and the long orbital period indicate that this is a pulsar-white dwarf binary.  Figure~\ref{phaseogram_1} (top right) shows the photon phaseogram and folded gamma-ray profile phase-aligned with the 327~MHz radio profile. This pulsar is very bright in gamma rays and the pulsations were detected with an H-test of 1267.99, corresponding to a significance of $\sim 34\sigma$.
%$34.21$\sigma$.

%Read 43268 events
%htest = 1255.60, significance = 34.03 sigma

\subsection{PSR~J1805$+$0615}

This black widow MSP was discovered in the \emph{Fermi}-LAT source 3FGL~J1805.9+0614 (4FGL~J1805.6+0615). The radio timing residuals vs.~observing epoch are shown in Figure~\ref{fig_res}. The \emph{Fermi} timing solution is shown in Table~\ref{tab:bw}. The pulsar has an orbital period of 8.08 hours and a minimum companion mass of 0.023~\msun. Figure~\ref{phaseogram_1} (bottom left) shows the photon phaseogram and folded gamma-ray profile phase-aligned with the 327~MHz radio profile. Gamma-ray pulsations were detected with a H-test of 183.29 $\sim 12\sigma$.
%(12.06$\sigma$). 

%Read 72290 events
%htest = 182.71, significance = 12.04 sigma

\subsection{PSR~J1824$+$1014}
This MSP was discovered in the \emph{Fermi}-LAT source 3FGL~J1824.0+1017 (4FGL~J1824.1+1013). The radio timing residuals vs.~observing epoch are shown in Figure~\ref{fig_res}. The \emph{Fermi} timing solution is shown in Table~\ref{tab:wd}. The pulsar has an orbital period of 82.5 days, the longest presented in this work. While long-$P_{\rm b}$ binaries appear to be rare among those found in \emph{Fermi}-LAT unidentified sources, the opposite is true for MSPs found in untargeted pulsar surveys; Section~\ref{sec:discussion} discusses this discrepancy in more detail. This pulsar's minimum companion mass is 0.27~\msun, indicating a white dwarf companion. Figure~\ref{phaseogram_1} (bottom right) shows the photon phaseogram and folded gamma-ray profile phase-aligned with the 327~MHz radio profile. Gamma-ray pulsations are weak and were detected with a H-test of 52.55 (6.15$\sigma$). 

%Read 79348 events
%htest = 52.55, significance = 6.15 sigma

\subsection{PSR~J1908$+$2105}

This eclipsing MSP was discovered in a source from a preliminary version of the 3FGL \emph{Fermi}-LAT catalog (P7R4J1909+2102). The source was not included in the final version of 3FGL, but is included in the newer 4FGL \emph{Fermi}-LAT source catalog as 4FGL~J1908.9+2103. The radio timing residuals vs.~observing epoch and orbital phase are shown in Figures~\ref{fig_res} and \ref{fig_resorb}. The \emph{Fermi} timing solution is shown in Table~\ref{tab:rb}. The pulsar has an orbital period of 3.51 hours and a minimum companion mass of 0.055~\msun. While the short orbital period and relatively small companion mass { are typical for black widows, its extended eclipses are typical for redbacks} (see Section~\ref{sec:discussion} for a more detailed discussion). Figure~\ref{phaseogram_2} (top left) shows the photon phaseogram and folded gamma-ray profile phase-aligned with the 327~MHz radio profile. Gamma-ray pulsations were detected with a H-test of 374.66, corresponding to a significance of $\sim 18\sigma$.
%17.78$\sigma$. 

%Read 134203 events
%htest = 374.66, significance = 17.78 sigma

\subsection{PSR~J2006$+$0148}

This MSP was discovered in the \emph{Fermi}-LAT source 3FGL~J2006.6+0150 (4FGL~J2006.4+0147). The radio timing residuals vs.~observing epoch are shown in Figure~\ref{fig_res}. The \emph{Fermi} timing solution is shown in Table~\ref{tab:wd}. The pulsar has an orbital period of 15.63 hours and a minimum companion mass of 0.15~\msun, indicating a white dwarf companion. Figure~\ref{phaseogram_2} (top right) shows the photon phaseogram and folded gamma-ray profile phase-aligned with the 327~MHz radio profile. Gamma-ray pulsations were detected with a H-test of 216.40 (13.20$\sigma$). 

%Read 43167 events
%htest = 216.40, significance = 13.20 sigma

\subsection{PSR~J2052$+$1219}
%\cite{Cromartie16}

The pulsar was discovered in the \emph{Fermi}-LAT source 3FGL~J2052.7+1217 (4FGL~J2052.7+1218). The radio timing residuals vs.~epoch and orbital phase are shown in Figures~\ref{fig_res} and \ref{fig_resorb}, and the \emph{Fermi} timing solution is shown in Table~\ref{tab:bw}. This black widow binary has an orbital period of 2.75~h and a minimum companion mass of 0.034~\msun. The pulsar exhibits radio eclipses and variable delays in the radio pulse arrival time near ingress and egress (Figure~\ref{fig_2052_ecl}). The relatively abrupt ingress indicates a sharp leading edge of the companion. In contrast, during the first few minutes after every observed egress we see pulse delays of up to 0.25 in rotational phase. This indicates the pulses experience an extra dispersive delay while passing through a cloud of ionized gas that trails the companion. After optimizing the timing solution over the full span of the \emph{Fermi} mission, gamma-ray pulsations were detected with an H-test significance of 254.67, corresponding to $14.42\sigma$. The phaseogram is presented in Figure~\ref{phaseogram_2} (bottom left).

%Read 36238 events
%htest = 254.67, significance = 14.42 sigma

\cite{Zharikov19} detect an optical counterpart to PSR~J2052+1218 and estimate the companion to have a radius of $0.12-0.15~\rsun$, nearly filling its Roche lobe. An independent estimate of the distance to this system from the optical data is 3.94(12)~kpc, which is consistent with the less precise distance estimate of 3.9~kpc from the pulsar's DM and the YMW16 model of ionized gas in the Galaxy \citep{YMW16}. (Dispersion-based distances have typical uncertainties of up to 40\%.)

\subsection{PSR~J1048$+$2339}

An initial timing solution for this pulsar based on two years of Arecibo and GBT data was presented in \cite{Deneva16}. The precise timing position allowed the identification of a variable optical counterpart, an M4 main sequence star asymmetrically heated by the pulsar wind. The timing solution also uncovered rapid quasi-periodic oscillations in the orbital period likely due to changes in the gravitational quadrupole moment of the companion. Because of these oscillations it was not feasible to extrapolate the pulsar ephemeris outside the range of radio observations. Analyzing \emph{Fermi}-LAT data within its 2-year range did not yield a detection of gamma-ray pulsations at the time. In Table~\ref{tab:rb} we present an updated timing solution based on 2.9 years of radio data that includes three additional orbital period derivatives. Figures~\ref{fig_res} and \ref{fig_resorb} show the radio timing residuals vs.~observing epoch and orbital phase. A detection of gamma-ray pulsations in \emph{Fermi}-LAT data from the same time range is shown in Figure~\ref{phaseogram_2}. The weighted H-test significance of the detection is 84.95, which corresponds to $\sim 8\sigma$; the phaseogram is presented in Figure~\ref{phaseogram_2} (bottom right). 

%Read 5015 events
%htest = 84.95, significance = 7.97 sigma

\section{Discussion}\label{sec:discussion}

The present work continues the trend of pulsar searches in \emph{Fermi}-LAT unidendified sources discovering a disproportionate number of redback and black widow systems with orbital periods of a few hours. Many of these systems exhibit radio eclipses, as is the case for four of our eight MSPs (Figure~\ref{fig_resorb}). 

\subsection{Targeted vs.~Untargeted MSP Searches and Discoveries}

\begin{figure}
    \centering
    \includegraphics[width=0.9\textwidth]{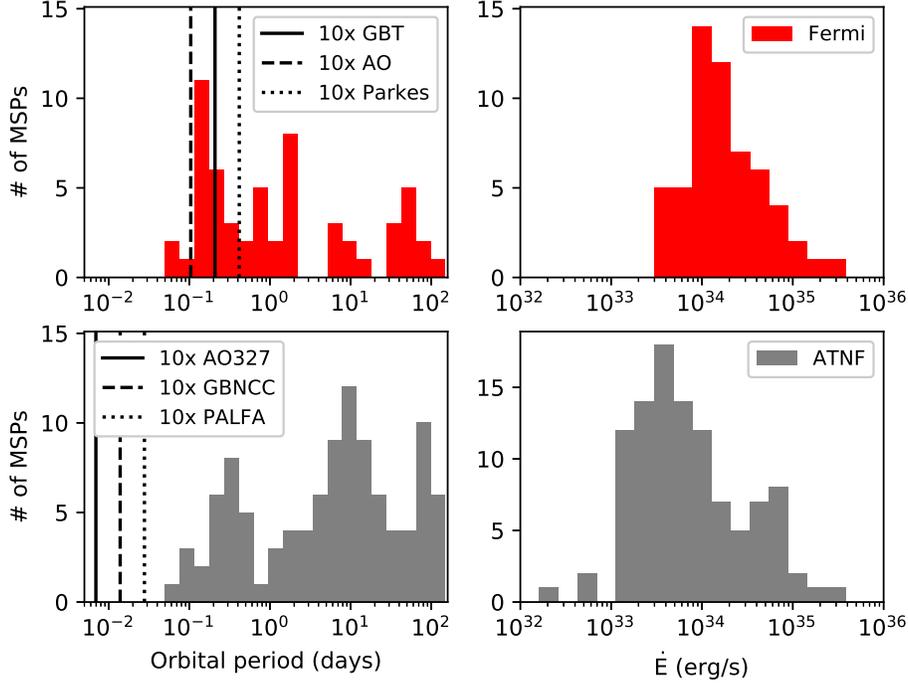}
    \caption{Top: Histograms of MSPs found in searches of \emph{Fermi}-LAT unidentified sources vs orbital period (left) and $\dot{E}$ (right). In the top left panel vertical lines show $10 \times T_{\rm obs}$ for searches of such sources at three observatories. Bottom: The same histograms for field MSPs from ATNF, excluding those found in \emph{Fermi}-LAT unidentified sources. In the bottom left panel vertical lines show $10 \times T_{\rm obs}$ for several untargeted radio pulsar surveys. Searches for binary pulsars assume constant line-of-sight acceleration; this assumption begins to break down for $P_{\rm b} < 10 \times T_{\rm obs}$.}
    \label{fig:pbhist}
\end{figure}

%A total of 22 black widows and 9 redbacks (including unpublished ones) 
A total of 18 black widows and 7 redbacks (not including unpublished ones) have been found since 2008 in radio searches of \emph{Fermi}-LAT unidentified sources. In contrast, there are 15 field black widows and 4 field redbacks in ATNF whose discovery was unrelated to \emph{Fermi} and took place over the five decades since the first binary pulsar was discovered. This raises the question of whether MSP gamma-ray emission is intrinsically correlated with a short orbital period or this apparent dependence is due to selection effects. Known gamma-loud MSPs have higher $\dot{E}$ on average than MSPs detected only in radio \citep{2PC}. We assembled two samples of binary MSPs whose $\dot{E}$ is known\footnote{Values of $\dot{E}$ used throughout this work are not corrected for the Shklovskii effect. This choice avoids heterogeneity in our comparison samples as many MSPs found in \emph{Fermi}-LAT unidentified sources do not have the proper motion measurements needed for the correction, including four of the eight in this work.} and have a rotation period $P < 10$~ms: those found in \emph{Fermi}-LAT unidentified sources, and all \emph{other} field MSPs in ATNF meeting these criteria. Figure~\ref{fig:pbhist} shows histograms of the two field MSP samples vs.~orbital period ($P_{\rm b}$) and $\dot{E}$. Both sets of panels show visibly different distributions. We perform Anderson-Darling tests on the $P_{\rm b}$ and $\dot{E}$ distributions of the two samples, confirming that they are significantly different (same-distribution null hypothesis probabilities $<0.001$, i.e. lower than the floor of probability values reported by {\tt scipy.stats.anderson\_ksamp}).
Kolmogorov-Smirnov tests on the $P_{\rm b}$ and $\dot{E}$ distributions of the two samples using {\tt scipy.stats.ks\_2samp} concur, yielding same-distribution null hypothesis probabilities of $2\times 10^{-5}$ and $1\times 10^{-6}$, respectively. 

MSPs are typically found in targeted or untargeted observations which are searched for pulsed signals assuming a range of constant line-of-sight accelerations of the pulsed source with respect to the telescope. The assumption of constant acceleration begins to break down for $P_{\rm b} \lesssim 10\times T_{\rm obs}$. We therefore overlay vertical lines on the two panels of Figure~\ref{fig:pbhist} corresponding to $10\times T_{\rm obs}$ of standard search observations that have yielded MSP discoveries in both of our samples. Targeted radio observations of \emph{Fermi}-LAT unidentified sources (top panels) tend to be $15-60$~minutes, while untargeted radio pulsar survey observations (bottom panels) tend to be $1-5$~minutes. We see that in the top left panel, MSP discoveries drop off drastically for $P_{\rm b}$ smaller than the shortest $10\times T_{\rm obs}$ of targeted search observations. This indicates that new discoveries are potentially acceleration-limited. It also suggests that such observations should be searched not only as one continuous $T_{\rm obs}$ but also separately in chunks of a shorter duration, down to 5~minutes, which would provide some continuity with untargeted surveys in acceleration-limited sensitivity. Another approach is to use jerk searches, which do not assume a constant line-of-sight acceleration but add its first derivative as an extra search dimension \citep{Tabassum20}.

In the bottom left panel in contrast, there is a gap showing no MSP discoveries over a range of $P_{\rm b} > 10\times T_{\rm obs}$. This indicates that untargeted surveys are likely flux-limited, in several different ways. First, the minimum detectable flux density, $S_{min} \propto 1/\sqrt{T_{\rm obs}}$, which puts shorter survey observations at a disadvantage compared to longer targeted ones. In untargeted surveys the center of the nearest radio beam would be offset from the (unknown) pulsar position by up to the half-power beam radius, which degrades $S_{min}$ by up to a factor of 2. Unidentified \emph{Fermi}-LAT sources are selected for targeted pulsar searches so that their error ellipses fit within the radio beam, and the beam center is placed at the error ellipse center, making any potential offset from the actual pulsar position likely much smaller than the beam radius.  

The spike of MSPs with short $P_{\rm b}$ in the top left panel of Figure~\ref{fig:pbhist} is matched by a decline in the bottom left panel. Untargeted surveys cover a position on the sky only once, while targeted searches of \emph{Fermi}-LAT unidentified sources observe every source of interest several times. This means that the latter are more likely to detect heavily eclipsing, short-$P_{\rm b}$ MSPs like J1048+2339 and J1908+2105 from the present work. On the other hand, a spike of MSPs with $P_{\rm b}$ on the order of a few days in the bottom left panel is matched by a scarcity of such discoveries in \emph{Fermi}-LAT sources in the top left panel. This can be explained by the fact that there is a historic bias in favor of long-$P_{\rm b}$ pulsars whose detectability is not significantly affected by line-of-sight acceleration. Many of these pulsars were easily detectable without and discovered before the advent of acceleration searches. A number of untargeted surveys from the past few decades with $T_{obs}$ ranging from a few minutes to an hour or longer have had a long head start in searching this part of the parameter space before targeted searches of \emph{Fermi}-LAT sources began. 

We conclude that selection effects are prominent enough that they cannot be ruled out as explanation for the different $P_{\rm b}$ distributions of MSPs found in \emph{Fermi}-LAT unidentified sources compared to MSPs found in untargeted all-sky surveys. Despite the different selection biases affecting our two MSP samples, there is a notable lack of discoveries with $P_{\rm b} \lesssim 1$~hour in both, even though some modern surveys using acceleration and jerk searches are in principle sensitive to signals in that part of the parameter space. In such cases targeted and untargeted radio pulsar searches alike may be hindered by an extreme manifestation of the effects making some redbacks { and black widows} difficult to detect: heavy eclipses, prolonged and uneven dispersive pulse delays like those shown in Figure~\ref{fig_2052_ecl}, and intense scattering by ionized gas in the system. { Dispersive and scattering effects would make even non-eclipsing pulsars with very short orbital periods difficult to detect.} Since gamma-ray emission is unaffected by dispersion and scattering, binary MSPs with $P_{\rm b} \lesssim 1$~hour may stand a better chance of detection when direct searches for gamma-ray pulsations from MSPs become computationally feasible. { This is exemplified by the two MSPs with the shortest orbital periods known, both of them black widow binaries: PSR~J1311$-$3430 ($P_b \, = \, 93$ min; \citealt{Pletsch12a}), which is very rarely detected at radio wavelengths, and PSR~J0653$-$0158 ($P_b \, = \, 75$ min; \citealt{Nieder20}), which has not been detected at radio wavelengths.}

\subsection{MSP Binary Types}

\begin{figure}
    \centering
    \includegraphics[width=0.9\textwidth]{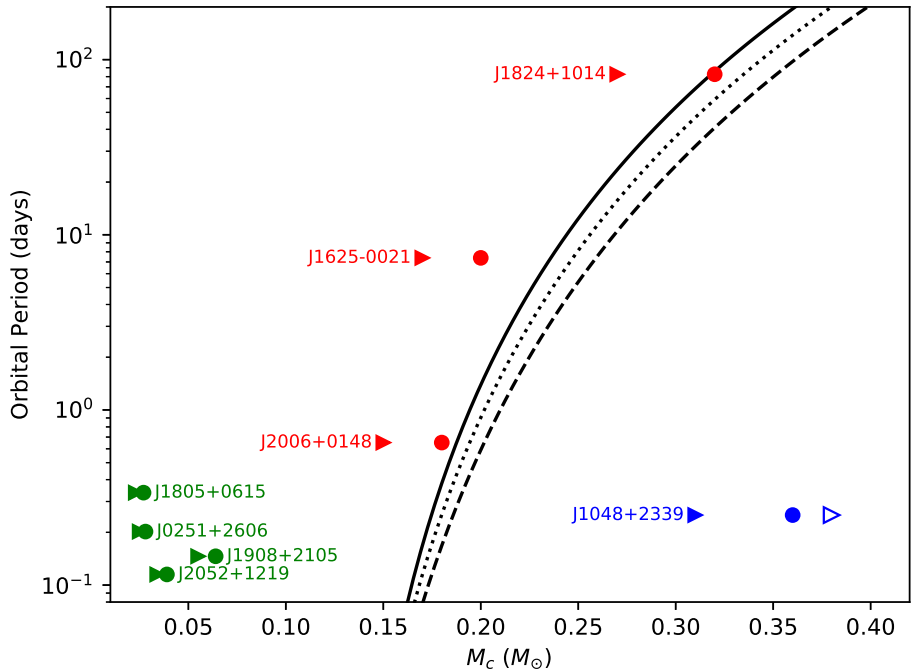}
    \caption{Curves show the $P_{\rm b}-M_c$ relation for the three populations of NS-WD binaries based on the metallicity of the WD progenitor considered by \cite{TS99}. Red triangles show the minimum companion masses for the three NS-WD systems in this work. Red circles show the companion mass for the median orbital inclination of $60\degree$. Green and blue symbols correspond to the black widow and redback systems in this work, respectively. { Filled symbols are based on pulsar timing in the present work and assume a pulsar mass of 1.4~\msun. The sole open triangle denotes a spectroscopic result from \cite{Strader19} for the companion mass of J1048+2339, corresponding to a pulsar mass of 1.96~\msun\ and inclination of $83\degrees$}.}
    \label{fig:pb-mc}
\end{figure}

\cite{TS99} derived a relation between the orbital period $P_{\rm b}$ and companion mass $M_c$ for three populations of NS-WD binary systems depending on the metallicity of the WD progenitor. Assuming a pulsar mass of 1.4~\msun\, Figure~\ref{fig:pb-mc} shows where our MSPs fall with respect to these relations based on their minimum companion masses (corresponding to an orbital inclination of $90\degree$) as well as the companion mass corresponding to the median orbital inclination of $60\degree$. Of the three NS-WD systems from Table~\ref{tab:wd} J1824+1014 and J2006+0148 conform well to the relation, while J1625$-$0021 does not. This indicates that the orbital plane of J1625$-$0021 may be less inclined than the other two systems: an inclination of $\sim 43\degrees-45\degrees$ corresponds to a $M_c = 0.24-0.25~\msun$, which would be consistent with the theoretical relation. Another possibility is that the neutron star in the J1625$-$0021 system is more massive than the assumed pulsar mass of 1.4~\msun. For the median orbital inclination of $60\degrees$, a pulsar mass of $1.9-2.1~\msun$ would also result in this system conforming to the relation. 

Redbacks fall in the lower right corner and black widows in the lower left corner of Figure~\ref{fig:pb-mc}. \cite{Cromartie16} classified J1048+2339 and J1908+2105 as redbacks based on the fact that both pulsars eclipse for nearly half of their orbital periods (Figure~\ref{fig_resorb}). Heavy eclipses are typical for redback systems due to their similarly short orbital periods but larger companions compared to black widow systems. In both types of binaries the intense pulsar particle wind gives rise to a plume of ionized gas around the companion that further contributes to the eclipse duration. 

{ Figure~\ref{fig:pb-mc} shows $M_c$ for the redback J1048+2339 with the same assumed pulsar mass and two inclinations as for the other MSPs in this work. It also shows a spectroscopic result from \cite{Strader19}, who argue that the pulsar mass in this system is $\geq 1.96~\msun$, and $M_c \geq 0.38$~\msun. The inclination corresponding to these parameters is $\geq 83^{+7}_{-10}$ degrees. Therefore either the orbit must be close to edge-on, or the pulsar mass must be signifcantly higher than 2~\msun.}

{ Even though J1908+2105 was previously classified as a redback, its short orbital period and relatively small minimum $M_c$ place it in the lower left corner in Figure~\ref{fig:pb-mc} together with the black widows from Table~\ref{tab:bw}.} However, the fact that J1908+2105 eclipses for $\sim 40\%$ of its orbit is unusual for a black widow. When black widows eclipse, they do so for no more than $\sim 10-20\%$ of their orbital periods (e.g. \citealt{Polzin18}, \citealt{Polzin19}, \citealt{Crowter20}). This points either to an atypical companion size in the J1908+2105 system and/or an unusually dense and extended ionized gas halo around the companion. { Alternatively, J1908+2105 could represent a middle-ground case between the two observational classes of redbacks and black widows. \cite{Chen13} deduce from stellar evolution simulations that redbacks do not evolve into black widows, but the bimodal distribution of properties between these two binary pulsar types is due to how efficiently the pulsar wind irradiates the companion, which in turn depends on beaming geometry. }
%Alternatively, J1908+2105 could be a NS-WD system with a low orbital inclination and low $M_c$. A companion mass of $0.1~\msun$ would correspond to an inclination of $34\degree$. 

\begin{figure}
    \centering
    \includegraphics[width=0.9\textwidth]{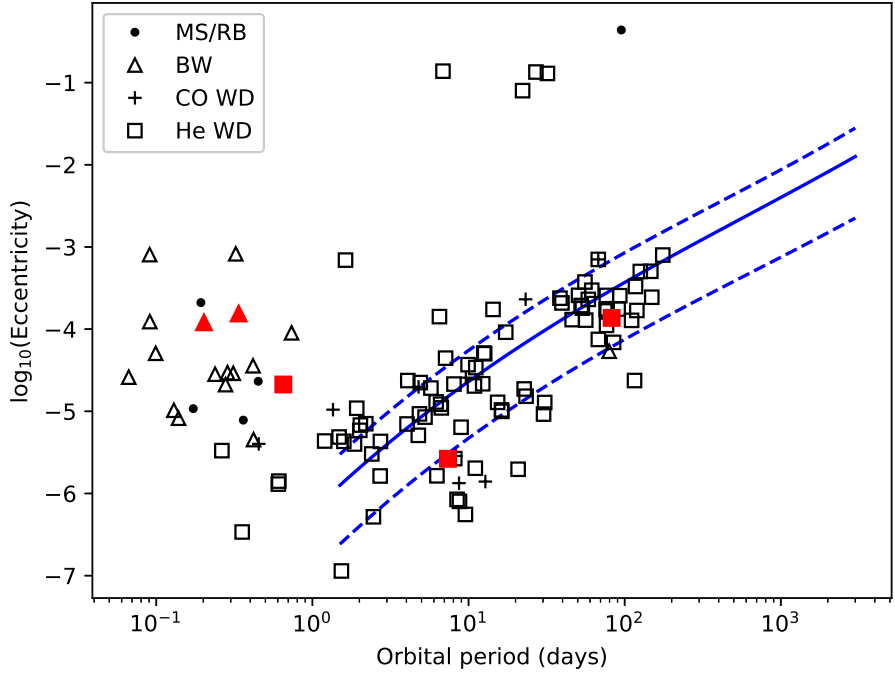}
    \caption{Curves show the theoretical relation between orbital period and eccentricity for NS-WD binaries (\citealt{Phinney92}, \citealt{PK94}). The solid curve corresponds to median values and the dashed curves correspond to 95\% confidence contours. Black symbols show field binary MSPs from the ATNF pulsar catalog as follows: redbacks or other systems with a main sequence companion (dots), black widows (open triangles), CO WD companion (crosses), and He WD companion (open squares). Red filled symbols show MSPs from this work with non-zero eccentricities in their best-fit timing solutions. Red filled squares denote the MSPs with He WD companions (J1625-0021, J1824+1014, J2006+0148), and red filled triangles denote the black widows J0251+2606 and J1805+0615.}
    \label{fig:pb-ecc}
\end{figure}

Another useful comparison is to see where our MSPs fall with respect to the theoretical relation between eccentricity and orbital period in NS-WD systems derived by \cite{Phinney92} and expanded by \cite{PK94}. In Figure~\ref{fig:pb-ecc} the median values and 95\%\ uncertainty of this relation are shown for a range of typical orbital periods. For a comparable companion mass, binaries with wider orbits would take longer to tidally circularize. However, \cite{Phinney92} and \cite{PK94} show that if tidal forces were the only or the dominant factor determining the final eccentricity in the system, most NS-WD binaries should have eccentricities indistinguishable from zero. The tidal circularization time scale is shorter than the giant evolution stage of the companion by several orders of magnitude; during this stage the companion fills its Roche lobe and mass transfer to the NS is ongoing. Instead, the effect of tides on orbital evolution is partly counteracted by eddies and density fluctuations within the convective layer of the giant, such that after mass transfer stops and the companion reaches its WD stage, the eccentricity remains ``frozen'' at a small but measurably non-zero value. 

Five of our eight MSPs have non-zero eccentricities in their best-fit timing solutions. Figure~\ref{fig:pb-ecc} shows that two of the NS-WD systems in Table~\ref{tab:wd} (J1625-0021 and J1824+1014) conform well to the relation between eccentricity and orbital period, while the third (J2006+0148) does not. The relation was derived assuming stable mass transfer from a giant star filling its Roche lobe. Such systems are expected to have $P_{\rm b} > 1$~day and $0.16~\msun \lesssim M_c \lesssim 0.45~\msun$ in their NS-WD stage. J1625$-$0021 and J1824+1014 have $P_{\rm b} = 7$ and 83 days, respectively, while J2006+0148 has $P_{\rm b} = 15.6$ hours. Of the three MSPs J1824+1014 has the highest minimum companion mass, $M_c > 0.27~\msun$, while J1625$-$0021 and J2006+0148 have $M_c > 0.17$ and 0.15~\msun, respectively. We conclude that the properties of J2006+0148 indicate a relatively short or intermittent mass transfer from a companion whose giant stage did not last long enough for the orbit to become very circularized.

\section{Conclusions}

{ We have presented the detection of gamma-ray pulsations from eight binary MSPs found in radio searches of \emph{Fermi}-LAT unidentified sources, 11-year timing solutions based on radio and gamma-ray data for seven of the MSPs, and an updated radio timing solution for the remaining pulsar. Our investigation of a potential correlation between the presence of gamma-ray emission and a short orbital period revealed that selection effects are prominent enough that they cannot be ruled out as the main cause. Based on this analysis we also conclude that the chance of new discoveries in both new and existing pulsar search data can be improved if observations longer than $5-10$ minutes are processed using jerk search codes and/or in chunks of that duration in order to minimize the degradation of search sensitivity due to unaccounted for line-of-sight acceleration. The orbital periods and inferred minimum companion masses of our eight MSPs indicate that three are NS-WD binaries, three are black widows with companions whose mass is on the order or $0.02-0.03~\msun$, and one is a redback with a companion whose mass is $\gtrsim 0.3~\msun$. The remaining MSP (J1908+2105) exhibits heavy radio eclipses reminiscent of redbacks but has a minimum companion mass of $0.055~\msun$, which would be either unusually high for a black widow or unusually low for a redback. This system may represent a rare middle-ground case between these two observational classes.}

\medskip
We thank David Nice for providing a numerical representation of the curves describing the relation between orbital period and eccentricity in Figure~\ref{fig:pb-ecc}. We also thank David Smith for advice about working with \emph{Fermi}-LAT data. 

The Arecibo Observatory is operated by the University of Central Florida, Ana G. Mendez-Universidad Metropolitana, and Yang Enterprises under a  cooperative agreement with the National Science Foundation (NSF; AST-1744119). 

The \emph{Fermi}-LAT Collaboration acknowledges generous ongoing support from a number of agencies and institutes that have supported both the development and the operation of the LAT as well as scientific data analysis. These include the National Aeronautics and Space Administration and the Department of Energy in the United States, the Commissariat \`a l'Energie Atomique and the Centre National de la Recherche Scientifique / Institut National de Physique Nucl\'eaire et de Physique des Particules in France, the Agenzia Spaziale Italiana and the Istituto Nazionale di Fisica Nucleare in Italy, the Ministry of Education, Culture, Sports, Science and Technology (MEXT), High Energy Accelerator Research Organization (KEK) and Japan Aerospace Exploration Agency (JAXA) in Japan, and the K.~A.~Wallenberg Foundation, the Swedish Research Council and the Swedish National Space Board in Sweden. Additional support for science analysis during the operations phase is gratefully acknowledged from the Istituto Nazionale di Astrofisica in Italy and the Centre National d'\'Etudes Spatiales in France. This work performed in part under DOE Contract DE-AC02-76SF00515.

Work at NRL was supported by the NASA \emph{Fermi} program. 

\software{PRESTO\footnote{\tt https://www.cv.nrao.edu/~sransom/presto} \citep{Ransom02}, TEMPO\footnote{\tt http://tempo.sourceforge.net}, 
\texttt{PINT}\footnote{\tt https://github.com/nanograv/PINT} \citep{Luo19}}
\texttt{Dracula}\footnote{\tt https://github.com/pfreire163/Dracula} \citep{fr2018}

\end{document}